\documentclass[journal]{IEEEtran}
\usepackage[utf8]{inputenc} 
\usepackage{graphicx}
\usepackage{subfigure}
\usepackage{amsmath}
\usepackage{multirow}
\usepackage[varg]{newtxmath}
\usepackage{booktabs}
\usepackage{amsmath}

\usepackage[psamsfonts]{amssymb}
\usepackage{amsxtra}
\usepackage{threeparttable}
\usepackage{romannum}
\usepackage{stfloats}
\usepackage{cite}
\usepackage{color, xcolor}
\usepackage{threeparttable}
\usepackage{ulem}
\usepackage{algorithmic}
\usepackage{algorithm}

\ifCLASSINFOpdf
\else
\fi
\begin{document}

%
\title{Multimodal Urban Sound Tagging with Spatiotemporal Context}


\author{Jisheng Bai,~\IEEEmembership{Student~Member,~IEEE,}
	Jianfeng Chen,~\IEEEmembership{Member,~IEEE,}
	and ~Mou Wang,~\IEEEmembership{Student~Member,~IEEE}
	\thanks{This work was supported by National Natural Science Foundation of China under Grant No.61501374 and No.62071383, and by NSFC-Zhejiang Joint Fund for the Integration of Industrialization and Information No.U1609204. \textit{(Corresponding author: Jianfeng Chen.)} }
	\thanks{J. Bai, J. Chen and M. Wang are with the School of Marine Science and Technology, Northwestern Polytechnical University, Xi’an, China. (e-mail:  baijs@mail.nwpu.edu.cn; chenjf@nwpu.edu.cn; wangmou21@mail.nwpu.edu.cn)}
}


\maketitle
\pagenumbering{arabic}


\begin{abstract} 
Noise pollution significantly affects our daily life and urban development.
Urban Sound Tagging (UST) has attracted much attention recently, which aims to analyze and monitor urban noise pollution.
One weakness of the previous UST studies is that the spatial and temporal context of sound signals, which contains complementary information about when and where the audio data was recorded, has not been investigated.
To address this problem, in this paper, we propose a multimodal UST system that deeply mines the audio and spatiotemporal context together.
In order to incorporate characteristics of different acoustic features, two sets of four spectrograms are first extracted as the inputs of residual neural networks.
Then, the spatiotemporal context is encoded and combined with acoustic features to explore the efficiency of multimodal learning for discriminating sound signals.
Moreover, a data filtering approach is adopted in text processing to further improve the performance of multi-modality.
We evaluate the proposed method on the UST challenge (task 5) of DCASE2020.
Experimental results demonstrate the effectiveness of the proposed method.
	
	
\end{abstract}




\IEEEpeerreviewmaketitle




\section{INTRODUCTION}
\label{sec:INTRODUCTION}

\IEEEPARstart{N}oise pollution significantly affects the daily life of urban residents and causes severe defects, such as sleep disruption, heart disease and hearing loss \cite{Bello2019sonyc},  \cite{stansfeld2003noise}. Therefore, it is essential to detect and monitor the distribution of urban noise for improving public health and living conditions \cite{cartwright2019sonyc}. Monitoring and assessment of urban noise has gained significant importance due to the threats it poses \cite{2013Novel}. Environmental sound recognition (ESR) has been one of the important methods for urban noise monitoring as it is responsible for discriminating different noise types.

ESR has long been a fundamental task of acoustic scene analysis, which has attracted much attention recently. ESR has been effectively used in different scenarios involving bird sound detection \cite{BaiWCCF19}, sound event localization and detection and sound tagging \cite{salamon2015unsupervised}.
It is more difficult than music and speech recognition, since the distributions of different environmental sounds vary dramatically and changes rapidly from time to time \cite{zhang2020learning}. This problem is particularly challenging in urban environments. In order to address this problem, a special branch of ESR, i.e., urban sound tagging (UST) is used.
UST aims to detect whether a particular kind of noise is present in a sound recording. It is a fundamental task of ESR for preventing noise pollution.

Previous methods of UST are all conducted in acoustic domain, leaving other modalities, such as text or image, far from explored \cite{adapa2019urban, Kim2019}.
Particularly, to our knowledge, we did not observe any multimodal UST system yet.
Specifically, in real-world scenarios, an urban sound usually occurs along with human activities over time and space.
The information about when and where the urban sound was recorded will be stored as \textit{spatiotemporal context} text. 
It provides discriminative and complementary information of the acoustic environment.
For example, a car horn is more probable to be heard in a busy street rather than inside a park, while the sounds of factory machines are unlikely to be heard around midnight.

A UST system faces many problems in real-life scenarios.  
Especially, finding distinguishable acoustic patterns is quite difficult in the complex urban environment.
Existing UST methods usually take one type of acoustic feature as input.
However, urban sounds contains different non-stationary features and changes in a wide range of frequencies.
Hence, it is difficult to distinguish various sounds using one type of acoustic feature.
Deep learning models, especially convolutional neural networks (CNNs), have been widely used as classifier for UST \cite{kong2019cross, Iqbal2020, bai2019multi}.
But as the depth of CNN increases, the problem of gradients vanishing occurs, which affects the performance of UST.
Moreover, in an urban environment, it is time-consuming and difficult to collect a lot of data for various types of urban sound.
Therefore, there is a lack of sufficient data for some urban sound classes in a UST dataset, where the data volumes of some classes are obviously different.

In this paper, we implement the multimodal UST system where we first extract the hidden representation of the UST audio signals using CNN, and the hidden representation of the spatiotemporal context text using recurrent neural networks (RNNs). 
Then, we combine the hidden representations for the final audio tagging. 
To fully mine the information of audio signals, we propose to integrate several non-stationary acoustic feature extraction methods. 
To prevent the gradient vanishing problem, we employ the residual network \cite{he2016deep}.
To address the real-world limitation that some types of urban sounds lack sufficient training data, we introduce a data augmentation method named mixup \cite{zhang2018deep}, which was originally proposed for environmental sound classification.
The main contributions of this paper are summarized as follows: 
\begin{itemize}
	\item Two sets of non-stationary acoustic feature are studied for UST, and we present the appropriate feature for tagging each type of urban sound.
	To our knowledge, HPSS and log-linear spectrograms are applied to UST for the first time.
	Moreover, an ensemble of CNNs and residual CNNs models is employed to boost the performance of UST.
	\item  The spatiotemporal context is introduced as a complementary modality for UST. 
	To fully utilize the spatiotemporal context, we have implemented fully connected (FC) and RNN based multimodal fusion strategies and an outlier filtering method.
	\item We present an extensive discrimination analysis of UST conducted on coarse-level classes and observe some interesting results from the experiments.
		Specifically, engine sound usually disrupts the classification of mechanical sounds and human voices interfere the detection of other urban sounds.

\end{itemize}
The rest of the paper is organized as follows.
Section \ref{sec:RELATED WORKS} introduces related works.
Section \ref{sec:PROPOSED METHOD} describes the proposed multimodal learning system for UST.
Section \ref{sec:EXPERIMENTS} provides the experimental settings.
Section \ref{sec:RESULTS AND DISCUSSION} presents the results.
Finally, Section \ref{sec:CONCLUSIONS} concludes this paper.
\section{RELATED WORKS}\label{sec:RELATED WORKS}
A conventional UST system usually consists of two components---an acoustic feature extractor and a classifier.
The early works in UST mostly focus on applying conventional classifiers, including Gaussian mixtures models \cite{mesaros2016tut} and support vector machines \cite{chu2009environmental}.
Recently, deep neural networks, such as CNNs and RNNs, have gained popularity as they outperformed some of the conventional classifiers.
The convolution operations in CNNs are able to catch local invariant features from the time-frequency representations of audio signals.
Experimental results show that the CNN models can achieve state-of-the-art performance in ESR tasks \cite{hershey2017cnn, Xu2017, vuegen2018weakly}.
However, when the depth of a CNN is increased, a well-known gradient vanishing problem occurs, which may lead to poor performance of UST in real-world scenarios.

A feature extractor aims to extract discriminative auditory patterns. 
Feature extractor can be divided into stationary and non-stationary methods \cite{cowling2003comparison}.
In real world scenarios, non-stationary acoustic features have been widely studied, since non-stationary sound occurs in abundance as compared to stationary sounds. 
Among the non-stationary acoustic features, Mel-frequency cepstral coefficients (MFCCs) are widely used and achieve leading results in ESR tasks \cite{heittola2013context,salamon2014dataset}. 
However, MFCCs are sensitive to noise and using MFCCs in urban environment is challenging \cite{cotton2011spectral}.
Therefore, a robust non-stationary feature extraction method needs to be developed for helping us understand characteristic of urban sound.

An effective non-stationary feature extraction method is to extract local auditory patterns from time-frequency representations.
Some representative features in ESR tasks contain short-time Fourier transform (STFT), harmonic and percussive separation (HPSS), and filterbanks based spectrograms.
Particularly, log-Mel spectrogram has been a major feature in DCASE challenges \cite{serizel2018large, kong2019cross}. 
Mel-scale spectrograms have reached state-of-the-art performance in different tasks \cite{salamon2017deep,cakir2017convolutional,akiyama2019multitask}.
However, considering the complexity of urban environment, one type of acoustic feature can not be used to tag various urban sounds.
Therefore, strategies of fusing multiple acoustic features have been developed to further utilize complementary information from sound signals.
For example, a CNN with mixed log-Mel and Gammatone spectrograms method was proposed for environment sound classification \cite{zhang2018deep}.
Log-Mel, STFT, harmonic spectrograms and MFCCs are integrated for UST and achieve outstanding performance \cite{bai2019multi}.

As it is known, the general topic of multimodal learning has a long research history. 
The performance of a specific task may be improved by incorporating knowledge from associated domains, which provides complimentary information for the task \cite{ngiam2011multimodal, srivastava2014multimodal, Zambellionline}. Multimodal deep learning is an emerging field which involves integration of knowledge from image and text by deep multimodal fusion \cite{ramachandram2017deep,liu2018active, jan2017artificial}.
Audio signals are fused with visual and physiological information for emotion recognition \cite{ringeval2013introducing}. 
A method of fusing audio, visual and textual modalities was proposed for sentiment analysis \cite{poria2016fusing}. 
Given multiple complementary modalities, fusing multimodal information is a critical problem.
Early fusion and late fusion strategies were proposed to address this problem.
In early fusion strategies, all modalities are used to train a single model, but this can be rather challenging due to information redundancy \cite{ramachandram2017deep}. In late fusion strategies, deep representations of each single modality are merged as joint representations to train a classifier \cite{simonyan2014two,yang2017eeg}.
Late fusion strategies merge the multimodal information effectively and have become the main stream, but fusing audio and text modalities in UST has not been studied.\\
\section{PROPOSED METHOD}\label{sec:PROPOSED METHOD}

In this section, we first present the architecture of the proposed system in Section \ref{subsec:overview}, and then present the components of the system in Sections \ref{subsec:Feature} to \ref{subsec:Spatiotemporal}.\\

\subsection{System overview}\label{subsec:overview}
Given an urban sound recording $x$ with its time and location recorded as a $D_{s}$-dimension spatiotemporal vector $s$.
UST aims to partition $x$ into one or more of $C$ sound classes:
\begin{equation}
z= f\left(  x \vert \theta \right),
\end{equation}
where $ z \in  \mathbb{R}^{C}$ is the probability of $x$ belonging to the $C$ classes, $f(\cdot)$ is a classifier, and $\theta$ is the parameter of the classifier.

\begin{figure*}[htb]
	\centering
	\includegraphics[width=16.8cm]{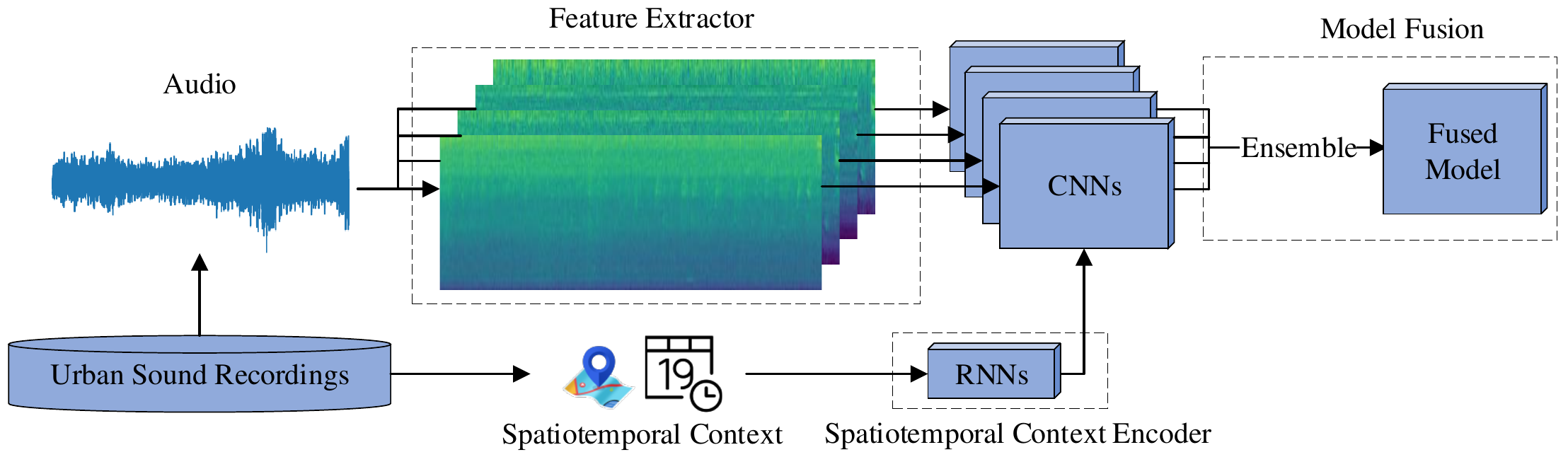}
	\caption{The overall architecture of the proposed multimodal UST system.}
	\label{fig:system}
\end{figure*}

The overall architecture of the proposed multimodal UST system is described in Fig. \ref{fig:system}.
It consists of four modules: a feature extractor, a spatiotemporal context encoder, a model fusion block and CNN models.
The feature extractor transforms $x$ into a time-frequency matrix $\mathbf{X} \in \mathbb{R}^{T\times F}$ of $T$ frames and $F$ frequency bins. 
The CNN architecture is trained to learn a nonlinear mapping function $g$:
\begin{equation}
\mathbf{O} = g\left(  \mathbf{X} \vert \theta_{1} \right),
\end{equation}
where $\mathbf{O}\in \mathbb{R}^{T' \times F' \times M}$ is acoustic representation with dimensions $T'$, $F'$ and $M$. $\theta_{1}$ is the parameter of CNN. Meanwhile, the spatiotemporal vector $s$ is encoded by RNN:
\begin{equation}
s' = h\left(  s \vert \theta_{2} \right),
\end{equation}
where $s' \in \mathbb{R}^{D_{s'}}$ is an encoded spatiotemporal vector of $D_{s'}$ dimension, $h(\cdot)$ represents the encoding function and $\theta_{2}$ is the parameter of the RNN.
We concatenate the audio and text representations as the input of the multimodal classifier to obtain the probability output $z$:
\begin{equation}
z = c\left(\mathbf{O}, s' \vert \theta_{3} \right),
\end{equation}
where function $c(\cdot)$ with its parameter $\theta_{3}$ is trained to classify urban sound recordings.

Finally, a hard decision will be made by comparing the probabilities to a threshold in $z$. The elements of $z$ that are higher than the threshold indicate the classification of $x$. It is noted that if there are several elements larger than the threshold, it indicates that several sound events occur simultaneously in $x$.

\subsection{Feature Extractor}\label{subsec:Feature}
This section describes two sets of acoustic features extracted by feature extractor.
The first set are Mel and linear filterbanks based spectrograms.
Log-Mel spectrogram is designed according to the speech perception of human auditory, and its resolution in low frequencies is relatively high \cite{kathania2019role}.
Log-Mel spectrogram has been manifested as one of the dominant features in speech processing and sound tagging \cite{huzaifah2017comparison}.
However, compared with speech signal, the frequency characteristics of urban sound can be more complex over the frequencies.
To investigate whether a higher resolution in high frequencies is helpful for UST, we propose to add linear filterbanks as a complement to the Mel filterbanks.

We can get a spectrogram $\mathbf{X}$ of an urban sound signal $x$ by applying STFT. An $A$-filterbank based spectrogram $\mathbf{Y} \in \mathbb{R}^{T \times A}$ is computed on $\mathbf{X}$:
\begin{equation}\label{equ:mel_spec}
	Y_{t,a}=\sum_{k=0}^{F-1} B_{k,a} \cdot X_{t,k}, \quad \forall a = 0,\ldots, A-1,
\end{equation}
where $Y_{t,a}$ is the time-frequency unit at the $t$th frame and $a$th filter band of $\mathbf{Y}$, and $B_{k,a}$ are triangular shaped filterbanks.
If $B_{k,a}$ in Eq. \ref{equ:mel_spec} are Mel filterbanks, we can obtain Mel-scale spectrogram and similarly for linear spectrogram.
Finally, we calculate log-Mel and log-linear spectrograms by applying logarithm to the obtained spectrograms.

The second set of acoustic features are Mel-scale HPSS spectrograms, including harmonic and percussive spectrograms.
Harmonic and percussive spectrograms contains different components separated from a sound signal.
By using these components, sound signals can be analyzed from different aspects.
For example, the harmonic components from music sounds and percussive components from impact sounds are more discriminative for classification.

Let $\mathbf{W}=\mathbf{X}^{2}$ be the power spectrogram, and $W_{t,k}$ and $X_{t,k}$ be the element of $\mathbf{W}$ and $\mathbf{X}$, respectively.
The harmonic power spectrogram $\mathbf{H}$ and the percussive power spectrogram $\mathbf{P}$ can be separated from $\mathbf{W}$ using HPSS algorithm \cite{fitzgerald2010harmonic,tachibana2013singing}. 
Similarly, we define $H_{t,k}$ as an element of  $\mathbf{H}$ and $P_{t,k}$ as an element of $\mathbf{P}$.
HPSS is formulated to find the optimal $\mathbf{H}$ and $\mathbf{P}$ by minimizing the function:
\begin{equation}
	\begin{aligned}
		J(\mathbf{H}, \mathbf{P})=& \frac{1}{2 \sigma_{H}^{2}} \sum_{t, k}\left(H_{t-1, k}-H_{t, k}\right)^{2} \\
		&+\frac{1}{2 \sigma_{P}^{2}} \sum_{t, k}\left(P_{t-1, k}-P_{t, k}\right)^{2}\\
		\text { subject to } &H_{t, k} \geq 0, P_{t, k} \geq 0 ,  \\
		\quad &  W_{t,k} = H_{t,k}+P_{t,k},\quad \forall(t, k)
	\end{aligned}
\end{equation}
where $\sigma_{H}^{2}$ and $\sigma_{P}^{2}$ are weighting parameters.
To take advantages of Mel-scale, $\mathbf{H}$ and $\mathbf{P}$ are filtered by Mel filterbanks and then their log values are computed.
In this paper, the Mel-scale harmonic and percussive power spectrograms are denoted as HPSS-h and HPSS-p, respectively.

\subsection{CNN model}\label{subsec:CNN}

The CNN model takes the acoustic features in Section \ref{subsec:Feature} as input. 
It consists of stacked convolutional blocks, each of which contains two convolutional layers, and the numbers of convolutional filters are the same inside each block.
Batch normalization (BN) \cite{ioffe2015batch} is applied to speed up the training process and prevent overfitting. Leaky rectified linear unit (ReLU) \cite{xu2015empirical} is used as non-linear activation.
After each convolutional block, an average pooling layer is added to reduce the feature maps.
The output of the last pooling layer is the high-level acoustic representation
$\mathbf{O}$ with dimensions of $T'$ (reduced frames), $F'$ (reduced frequency bins) and $M$ (the number of feature maps).
Then the mean values across the $F'$ dimension are computed, in the form of a matrix $\mathbf{O}' \in \mathbb{R}^{T' \times M}$.
TimeDistributed layers and AutoPool \cite{8434391} layer are used for producing the final prediction $z$.

To prevent the gradient vanishing problem of the CNN model, we replace the last convolutional block by a residual block.
Suppose the input of the residual block is $x'$, the original output before the second activation function is $a(x')$, the output after a convolutional layer with kernel size of $1\times1$ and BN is $b(x')$, and the final output is $y'$: 
\begin{equation}
y'=\sigma(a(x')+b(x')),
\end{equation}
where $\sigma$ denotes the Leaky ReLU activation function.

\begin{figure}[t]
	\begin{center}
		\includegraphics[width=8cm,height=9.4cm]{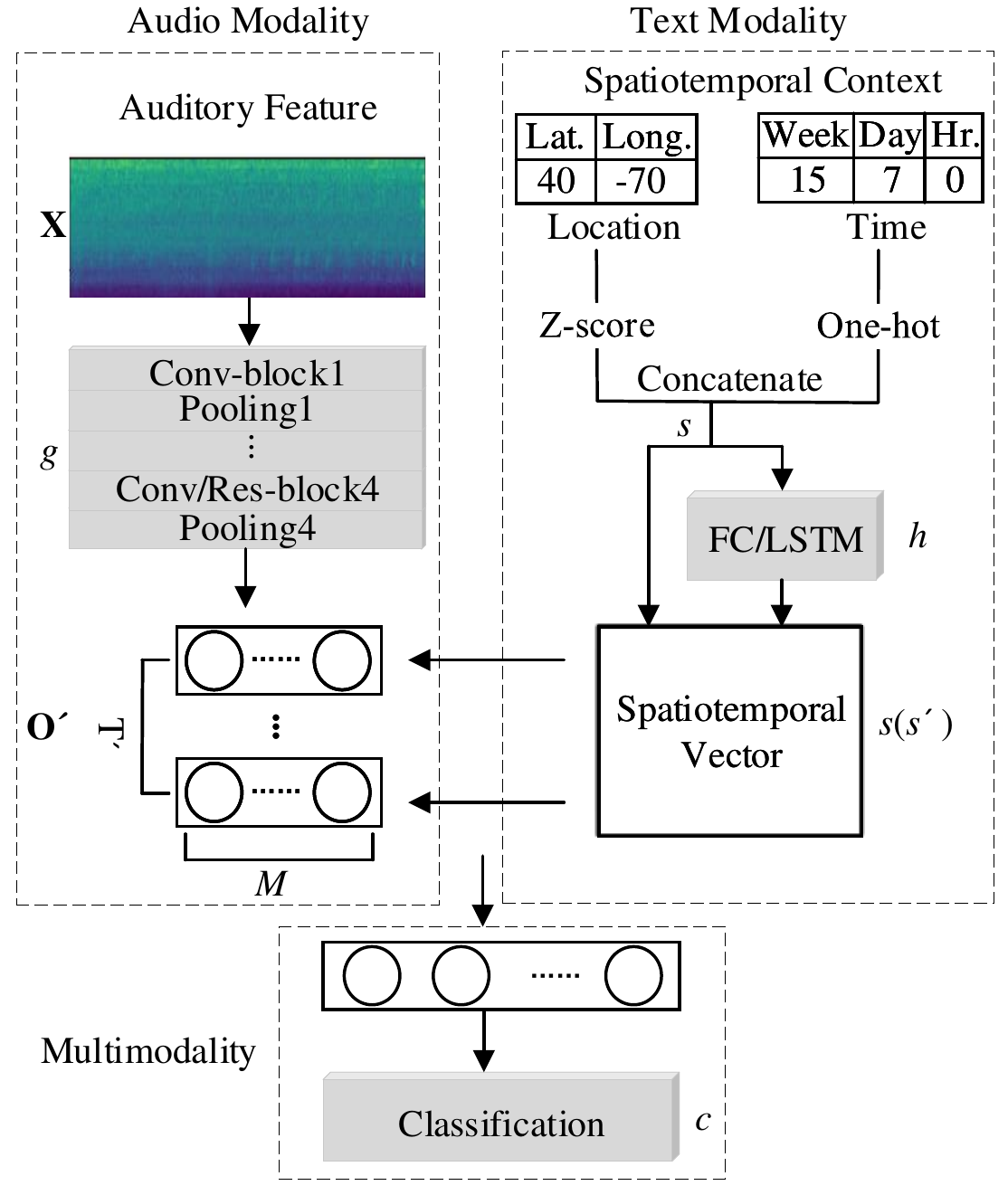}
	\end{center}
	\caption{The processing pipelines of multi-modality.}
	\vspace*{-3pt}\label{fig:spatiotemporal context}
\end{figure}

\subsection{Spatiotemporal Context Encoder}\label{subsec:Spatiotemporal}

As shown in Fig. \ref{fig:spatiotemporal context}, the spatiotemporal context of a recording consists of spatial context and temporal context.
The spatial context records the location where a sound occurs, in terms of latitude and longitude.
The temporal context records the time when a sound occurs, in terms of hour (24 hours in a day), day (7 days in a week), and week (52 weeks in a year).

Since the audio modality and text modality in the audio tagging problem have different statistical characteristics and dimensions, it is difficult to fuse the two modalities. Here we propose a multimodal processing pipeline as described in Fig. \ref{fig:spatiotemporal context}.
In the figure, audio modality extracts high-level acoustic representations from the acoustic features as in Section \ref{subsec:CNN}.
The text modality first normalizes the latitude and longitude values of the spatial context by the Z-score normalization, and transforms the temporal context to a one-hot vector. Then, it combines the spatial and temporal vectors into an 85-dimensional spatiotemporal vector ${s}$. 
An additional spatiotemporal vector $s'$ is obtained by encoding ${s}$ via a single layer, which can be FC or Long Short-Term Memory (LSTM) layer.
Below these parts, a multi-modality block fuses the audio and text modality for classification.
The output feature $\mathbf{O}'$ is concatenated with $s$ or encoded vector $s'$ for each frame. The fused multimodal feature is flattened and fed into the classification layer.

The multimodal learning algorithm is described in Algorithm 1.

\begin{algorithm}[t]
	\caption{Proposed multimodal learning method}
	\label{alg:multimodal}
	\begin{algorithmic}[1]
		\REQUIRE $N_{1}$ pairs of train spectrograms $\mathbf{X}_{t} = (\mathbf{X}_{1}, \cdots, \mathbf{X}_{N_{1}})$, spatiotemporal vectors $S_{t} = (s_{1}, \cdots, s_{N_{1}})$ and labels
		$L_{t}=(l_{1}, \cdots, l_{N_{1}})$,\\
		$N_{2}$ pairs of validate spectrograms $\mathbf{X}_{v} = (\mathbf{X}_{1}, \cdots, \mathbf{X}_{N_{2}})$, spatiotemporal vectors $S_{v} = (s_{1}, \cdots, s_{N_{2}})$ and labels $L_{v}=(l_{1}, \cdots, l_{N_{2}})$.
		\ENSURE  Predictions of validate set $\mathbf{Z}_{v} \in  \mathbb{R}^{C \times N_{2}}$.
		\STATE {Initialize parameters in CNN9 or CNN9-Res.}
		\REPEAT
		\STATE{\textbf{if} spatiotemporal vectors are not encoded \textbf{do}}
		\STATE  $\quad$Obtain $\mathbf{Z}_{t}\in \mathbb{R}^{C \times N_{1}}$ by $\mathbf{Z}_{t}= c(g(\mathbf{X}_{t}),S_{t} \vert \theta_{1},\theta_{3})  $;
		\STATE{\textbf{else}}
		\STATE $\quad$Obtain  $S'_{t}$ by $S'_{t} = h(S_{t} \vert \theta_{2})$;
		\STATE  $\quad$Obtain $\mathbf{Z}_{t}$ by $\mathbf{Z}_{t}= c(g(\mathbf{X}_{t}),S'_{t} \vert \theta_{1},\theta_{3})$;
		\STATE Update parameters $\theta_{1}$,$\theta_{2}$ and $\theta_{3}$ with $L_{t}$ and $\mathbf{Z}_{t}$;
		\STATE Obtain $\mathbf{Z}_{v}$ by using $\mathbf{X}_{v}$ and $S_{v}$ in step 3-7;
		\STATE Calculate evaluation metric with $L_{v}$ and $\mathbf{Z}_{v}$;
		\UNTIL{The evaluation metric does not improve in three steps.}
	\end{algorithmic}
\end{algorithm}

\subsection{Model Fusion}\label{subsec:Fusion}
An ensemble of CNN models can lead to better performance than its components \cite{wang2019}.
Thus, to take advantages of different acoustic representations for recognizing urban sound, we exploit a masking matrix $\mathbf{I}(\cdot)$  for model fusion, which can be expressed as follows:

\begin{equation}\label{equ:Fusion}
\mathbf{Z}=\sum_{u=1}^{U} \mathbf{Z}_{v}(u) \odot \mathbf{I}(u),
\end{equation}
where $\mathbf{Z} \in \mathbb{R}^{C \times N_{2} }$ is the prediction matrix of ensemble models, $U$ denotes the number of models, and $ \odot$ represents hadamard product. $\mathbf{I}(\cdot)$ has the same shape as  $\mathbf{Z}_{v}$ and is initialized as zero matrix. Its target column is set to 1 if the model achieves the best score on target class.

\section{EXPERIMENTS}\label{sec:EXPERIMENTS}

\subsection{Dataset}
We conducted experiments on the development dataset of DCASE2020 task5, named Sounds of New York City Urban Sound Tagging (SONYC-UST-V2) \cite{cartwright2020sonyc}.
SONYC-UST-V2 consists of 8 coarse-level urban sound categories, which are further divided into 23 fine-level sound categories. 
In this paper, we used coarse-level labels for evaluation. 
The names of the coarse-level classes and their abbreviations are listed in Table \ref{tab:abbr.}, which were further assigned with the identifiers $\{1,2,\ldots,8\}$ respectively. 
It is noted that a sound recording may have multiple coarse-level labels.

\begin{figure}[t]
	\centering
	\includegraphics[width=8cm,height=6cm]{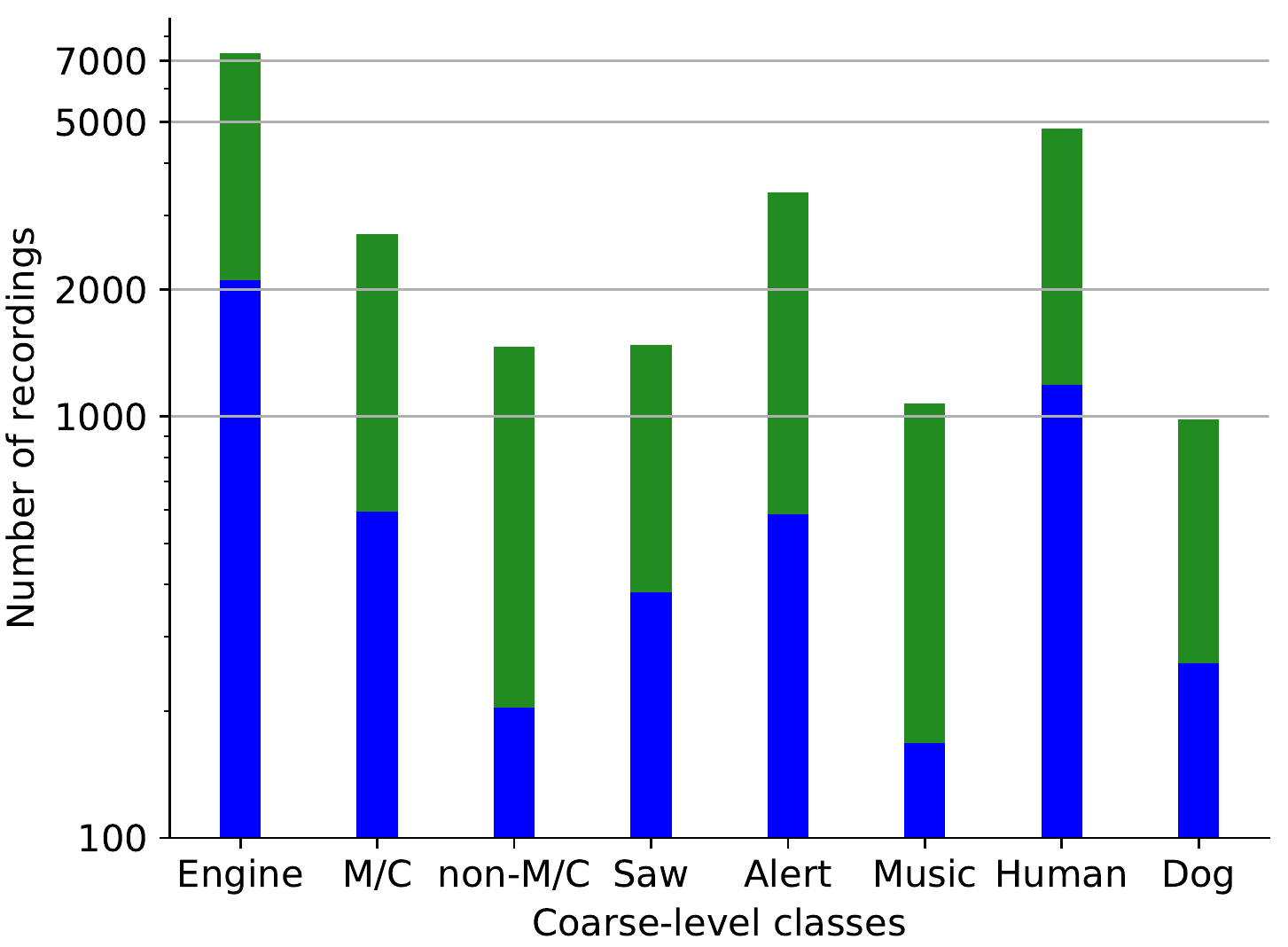}
	\caption{The number of urban sound recordings in each class. The bars in blue represent the recordings labeled with only one coarse-level category and the bars in green represent the recordings labeled with multiple coarse-level categories.}
	\label{fig:At least and only one}
\end{figure}

The numbers of recordings which only contain \textit{engine}, \textit{M/C}, \textit{Non-M/C}, \textit{Saw}, \textit{Alert}, \textit{Music}, \textit{Human}, or \textit{Dog} are 2110, 595, 204, 382, 585, 168, 1189 and 260, respectively.
Moreover, the numbers of recordings which contain \textit{engine}, \textit{M/C}, \textit{Non-M/C}, \textit{Saw}, \textit{Alert}, \textit{Music}, \textit{Human}, and \textit{Dog} are 7289, 2705, 1466, 1479, 3407, 1072, 4828 and 984, respectively.
Fig. \ref{fig:At least and only one} further demonstrates the distribution of recordings, where it can be seen that this is a class-imbalance problem. 

\begin{table}[t]
	\centering
	\setlength{\tabcolsep}{9mm}
	\renewcommand\arraystretch{1.25}
	\caption{Coarse-level classes and abbreviations}\label{tab:abbr.}
	\begin{tabular}{cc}
		\hline\hline
		Coarse-level classes       & Abbreviations \\ \hline
		Engine               & -       \\
		Machinery impact     & M/C           \\
		Non-machinery impact & Non-M/C       \\
		Powered saw          & Saw           \\
		Alert signal         & Alert         \\
		Music                & -         \\
		Human voice          & Human         \\
		Dog                  & -           \\ \hline\hline
	\end{tabular}
\end{table}

\subsection{Parameter Settings}
All urban sound recordings were resampled to a sampling rate of 22050 Hz. We extracted STFT spectrograms from the recordings using a Hanning window and hop length with the size of 1024 and 512 samples.
Mel filters of 64 bands were used to transform STFT and HPSS spectrograms to Mel-scale spectrograms.
Meanwhile, linear filters with bands of 64 were applied to obtain linear spectrograms.
Finally, the obtained spectrograms were transformed to decibels.

To summarize, the proposed CNN framework consists of 9 layers. 
We denote the CNN architecture as CNN9 and the residual architecture as CNN9-Res.
The detailed architecture and parameter settings of CNN9 are outlined in Table \ref{tab:para_nn}.

\begin{table}[t]
	\centering
	\setlength{\tabcolsep}{9.5mm}
	\renewcommand\arraystretch{1.25}
	\caption{The detailed architecture and parameter settings of CNN9.}
	\begin{threeparttable}
		\begin{tabular}{cc}
			\hline\hline
			Blocks   & Settings \\ \hline
			Conv-block1   &($3\times 3$\tnote{*}, 64\tnote{*}, BN, ReLU) $\times 2$\tnote{*}    \\
			Pooling1    & $2\times 2$\tnote{*} Average Pooling   \\
			Conv-block2    &($3\times 3$, 128\tnote{*}, BN, ReLU) $\times 2$  \\
			Pooling2    & $2\times 2$ Average Pooling   \\
			Conv-block3    &($3\times 3$, 256\tnote{*}, BN, ReLU) $\times 2$   \\
			Pooling3    & $2\times 2$ Average Pooling   \\
			Conv-block4    &($3\times 3$, 256, BN, ReLU) $\times 2$   \\
			Pooling4    &$1\times 1$ Average Pooling  \\
			Dense    & TimeDistributed       \\
			Dense    &TimeDistributed      \\
			AutoPooling & AutoPool1D           \\ \hline\hline
		\end{tabular}\label{tab:para_nn}
		\begin{tablenotes}
			\footnotesize
			\item[*] The terms $3\times 3$ and $2\times 2$ are the kernel sizes of the convolution and pooling layers respectively. The term $\times 2$ represents two  convolutional layers in each convolutional block. The numbers 64, 128 and 256 are the number of filters in the convolutional blocks.
		\end{tablenotes}
	\end{threeparttable}
\end{table}

For training, binary cross entropy was used as loss function, and batch size was set to 64. 
The Adam optimizer was adopted with a learning rate of 0.001. 
If the evaluation metric did not improve for successive three epochs, we stopped training. We recorded the training model at each epoch, and used the one that achieved the highest score for evaluation.

Data augmentation is an effective way to improve generalization and prevent overfitting of the neural networks.
In our system, we employ mixup as the data augmentation method in the training stage \cite{zhang2017mixup}. The mixup operations on the training samples are as follows:
\begin{equation}
\centering
\tilde{x}=\lambda x_{i}+(1-\lambda) x_{j}
\end{equation}
\begin{equation}
\tilde{y}=\lambda y_{i}+(1-\lambda) y_{j},
\end{equation}
where $x_{i}$ and $x_{j}$ are the input features, $y_{i}$ and $y_{j}$ are the corresponding target labels and $\lambda\in[0,1]$ is a random number drawn from the beta distribution.
As the new samples in the mixup operation are generated by linearly interpolating two real samples within a batch data, we can get more training samples without extra computing resource.
In addition to the one-hot labels, the spatiotemporal context in the next section are also mixed during training steps.

\subsection{Comparative Methods}
We compared our proposed method with the most recent approaches for UST in the following.
\subsubsection{Two-branch neural network (TBNN)}
This method \cite{Diez2020} is based on a multi-input architecture which consists of two different branches.
The audios are fragmented into 1 second frames, and log-Mel spectrograms with 128 Mel bands are used as input of a CNN branch.
The CNN uses two convolutional and max-pooling layers. 
Another branch consists of two FC layers for processing spatiotemporal context. 
The output of two branches are merged and fed to two FC layers.
\subsubsection{Baseline-2020}
This baseline system \cite{Cartwright2020} is provided by the organizers of DCASE2020 task5.
The baseline system uses a multi-label multi-layer perceptron model using a single hidden layer of size 128, and it is followed by an AutoPool layer.
The baseline system takes OpenL3 embeddings \cite{8682475} as sound representation, and one hot vectors as spatiotemporal context representation. 
\subsubsection{Randomly-initialized gated CNN (GCNN)}
The GCNN \cite{Iqbal2020} achieved state-of-the-art performance of DCASE2020 task5.
The GCNN takes log-Mel spectrogram as the input and contains ten gated convolutional (GC) layers \cite{dauphin2017language} with the following number of output feature maps: 64, 64, 128,
128, 256, 256, 512, 512, 512, 512. 
Following the GC layers, the time and frequency dimensions are reduced to a scalar using average pooling.
Spatiotemporal context is passed through a FC layer with 52 output features and then merged with the embeddings of the GCNN.
SpecAugment \cite{park2019specaugment} is used as a form of data augmentation during training.
\subsubsection{Pre-trained audio neural network (PANN)}
The PANN won the first place on the evaluation dataset of DCASE2020 task5.
The pre-trained CNN is the "CNN10" PANN proposed by Kong et al. \cite{kong2020panns}, which was pre-trained on AudioSet \cite{gemmeke2017audio}. 
It contains eight convolutional layers and two FC layers, and takes log-Mel spectrogram as the input. 
The processing of spatiotemporal context and data augmentation method are the same as GCNN.
Finally, a mean ensemble of four PANN models are used as the final model for UST.

\subsection{Evaluation Metrics}\label{subsec:Evaluation Metrics}
The area under the precision-recall curve (AUPRC) is a common evaluation metric for classification.
We calculate AUPRC to evaluate class-wise performance.
The calculation of AUPRC can be derived as follows.
Assuming equal weights are given to every sample, we can obtain true positives (TP), false positives (FP), and false negatives (FN) for each coarse-level category. 
Then we compute precision $P = TP/(TP+FP)$ and recall $R=TP/(TP+FN)$.
By varying the threshold $\tau$ between 0 and 1, different $P$ and $R$ will be obtained.
After that, these different $P$ and $R$ are used to draw the $P-R$ curve.
Finally, we compute the area under the $P-R$ curve to obtain the class-wise AUPRC.

The annotations of urban sound recordings in the UST system are labeled as multiple classes.
In order to evaluate the performance of the UST system, macro-averaged AUPRC is calculated on the coarse-level categories.
In this paper, we denote the macro-averaged AUPRC as macro-auprc.

\section{RESULTS AND DISCUSSION}\label{sec:RESULTS AND DISCUSSION}
In this section, experiments are conducted to evaluate the proposed multimodal UST system.
The results of comparative methods are reported in Table \ref{tab:competitors}.
	From the table, it is seen that the proposed method achieved a macro-auprc score of 0.775 on the DCASE2020 task5 validate dataset, which is greater than the previous methodologies.
The followings describe the detailed results and discussions.\\

\begin{table}[h]
		
		\centering
		\setlength{\tabcolsep}{9.6mm}
		\renewcommand\arraystretch{1.25}	
		\caption{Comparison with other methods in terms of macro-auprc on the DCASE2020 task5 validate dataset.}
			
		\begin{tabular}{cc}
			\hline\hline
			Methods & Macro-auprc  \\ \hline
			TBNN \cite{Diez2020} & 0.591\\
			Baseline-2020 \cite{Cartwright2020}   & 0.632       \\
			GCNN \cite{Iqbal2020} & 0.676\\
			PANN \cite{Iqbal2020} & 0.767\\ \hline	
			Our system & \textbf{0.775}   \\
			\hline\hline		
		\end{tabular}
		\label{tab:competitors}
		
\end{table}

\subsection{Multimodal Learning}

To study the effectiveness of the multimodal learning for UST, we first compared the proposed method with the method that does not include the spatiotemporal context, where we used CNN9 without mixup and took log-Mel spectrogram as the acoustic feature. 
We further evaluated the FC encoder and LSTM encoder, which are two different methods of encoding the spatiotemporal context.
The experimental results are given in Table \ref{tab:encoded mothod}.

\begin{table*}[h]
	\centering
	\renewcommand\arraystretch{1.25}
	\caption{Results of single, multiple modality and encoding methods for each coarse-level class.}
	\label{tab:encoded mothod}
	\begin{threeparttable}
	\setlength{\tabcolsep}{4.5mm}{	
		\begin{tabular}{cccccccccc}
			\hline\hline
			STC\tnote{*} vector & Modality & Engine & M/C   & Non-M/C & Saw   & Alert & Music & Human & Dog  \\ \hline
			-   & Single\tnote{**}    &	\textbf{0.882}&	0.622&	0.593&	0.743&	\textbf{0.955}&	\textbf{0.682}&	0.977&	0.480 \\		
			Original & Multiple\tnote{**} & 0.877  & 0.637 & 0.586   & \textbf{0.758} & 0.948 & \textbf{0.682} & 0.976 & 0.306   \\	
			Original filtered &Multiple & -  & - &\textbf{0.604} & - & - & - & - & \textbf{0.5}   \\
			FC encoded &Multiple  & 0.875  & 0.617 & 0.568   & 0.743 & 0.949 & 0.679 & 0.977 & 0.375    \\
			LSTM encoded & Multiple & 0.875  & \textbf{0.664} & 0.587   & \textbf{0.758} & 0.948 & 0.675 & \textbf{0.978} & 0.348  \\ \hline\hline
	\end{tabular}}
			
\begin{tablenotes}
	\footnotesize
	\item[*] The term STC is short for spatiotemporal context.
	\item[**] Single denotes the method of using audio modality. Multiple denotes the method of using audio and text modalities.
\end{tablenotes}
\end{threeparttable}
\end{table*}

\subsubsection{Single modality vs. Multi-modality}

The results with spatiotemporal context show improvement for some classes, such as \textit{M/C} and \textit{saw}, but worse results are obtained on other classes. 
To address this problem, we analyzed the location and time distribution of \textit{non-M/C}.

\begin{figure}[h]
	\centering
	\includegraphics[width=9cm,height=6cm]{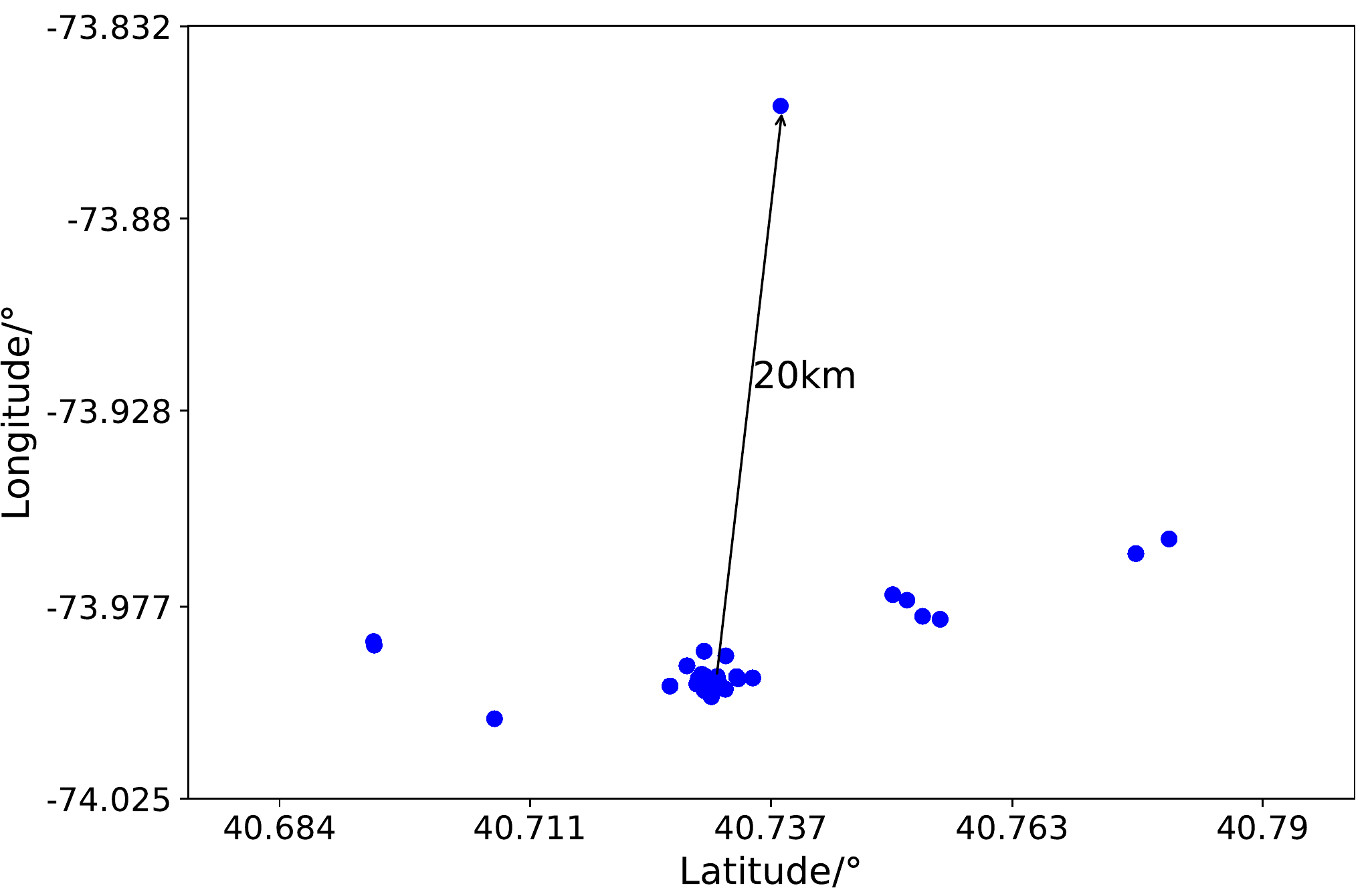}
	\caption{Location distribution of non-M/C}
	\label{fig:Lolazation for nonMC}
\end{figure}

\begin{figure*}[h]
	\centering
	\subfigure[Week distribution of non-M/C]{
		\begin{minipage}[b]{0.31\textwidth}
			\includegraphics[width=1\textwidth]{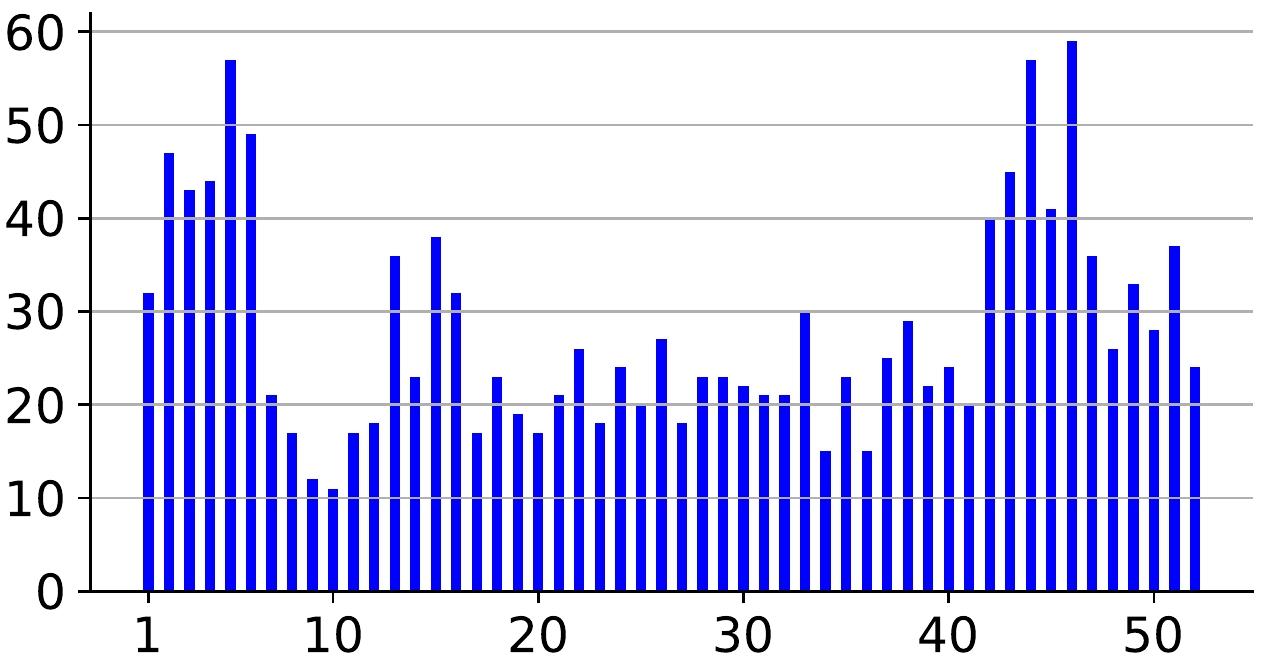}
		\end{minipage}\label{subfig:Time distribution1}
	}
	\subfigure[Day distribution of non-M/C]{
	\begin{minipage}[b]{0.31\textwidth}
		\includegraphics[width=1\textwidth]{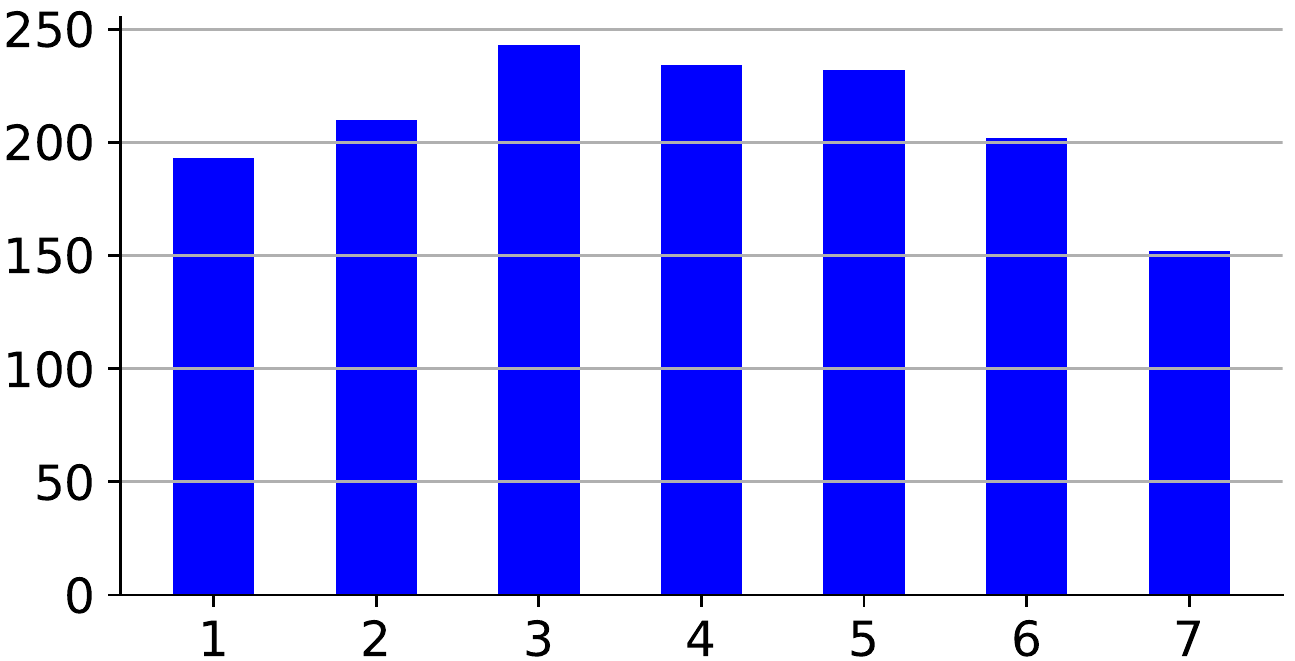}
	\end{minipage}\label{subfig:Time distribution2}
}
	\subfigure[Hour distribution of non-M/C]{
	\begin{minipage}[b]{0.31\textwidth}
		\includegraphics[width=1\textwidth]{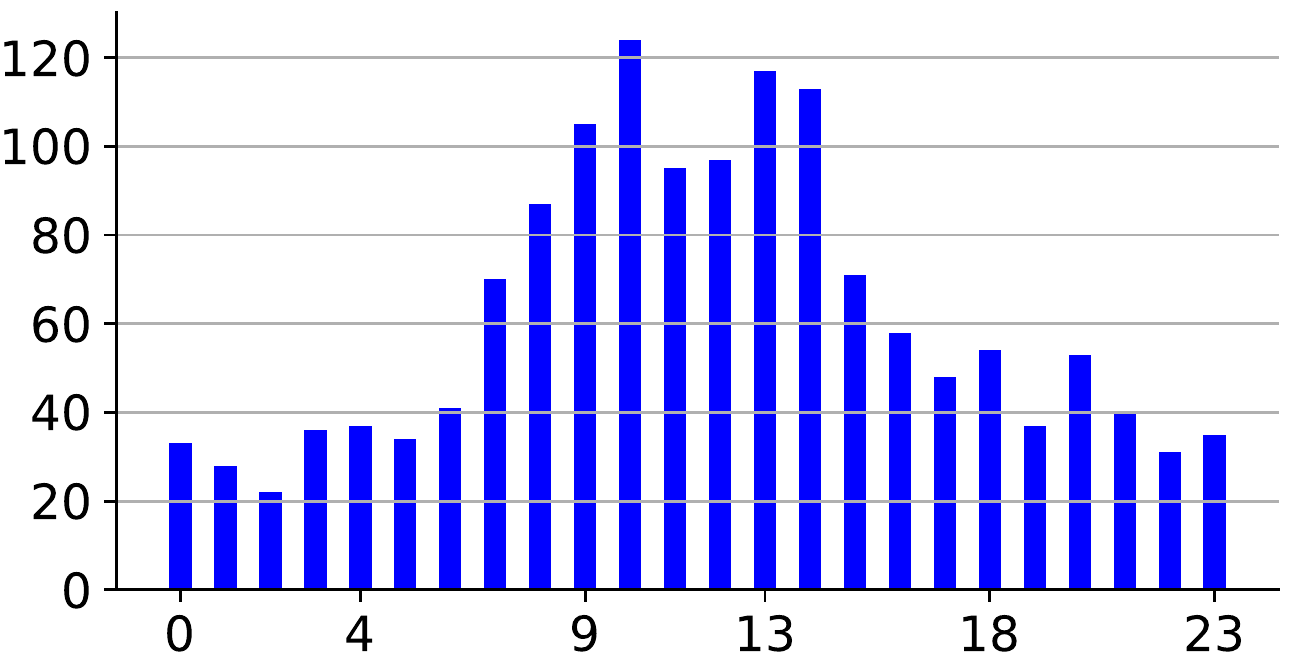}
	\end{minipage}\label{subfig:Time distribution3}
}
\\
	\subfigure[Filtered week distribution of non-M/C]{
		\begin{minipage}[b]{0.31\textwidth}
			\includegraphics[width=1\textwidth]{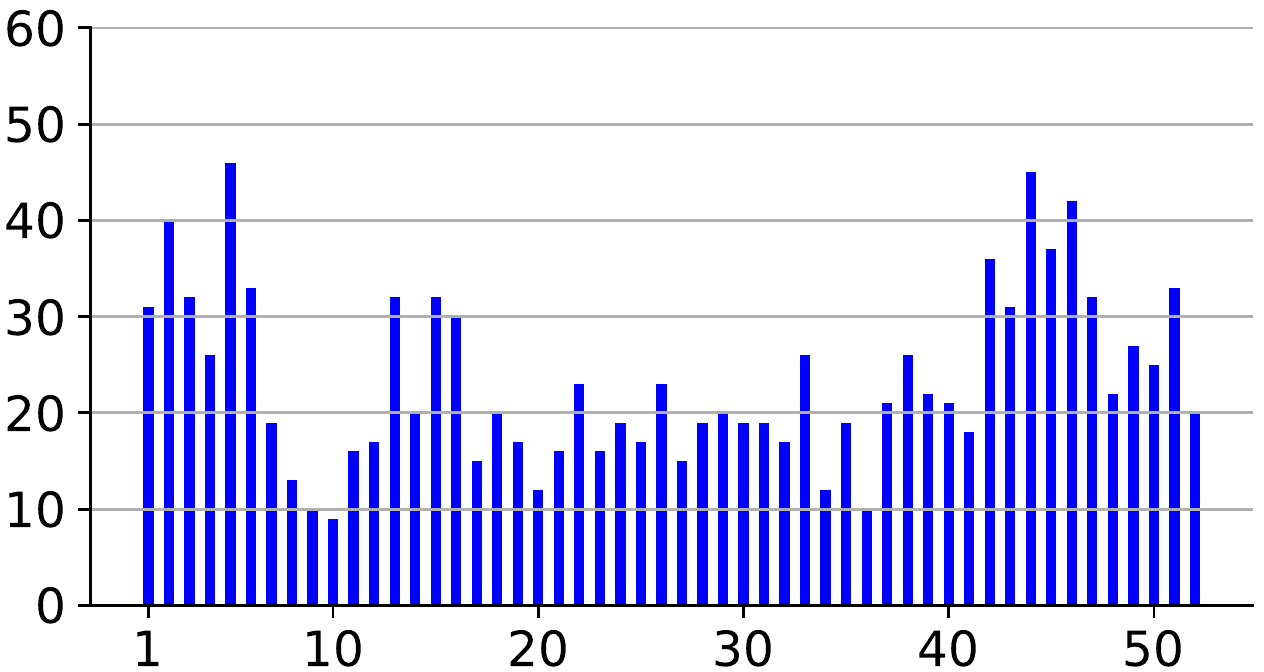}
		\end{minipage}\label{subfig:Time distribution4}
	}
	\subfigure[Filtered day distribution of non-M/C]{
		\begin{minipage}[b]{0.31\textwidth}
			\includegraphics[width=1\textwidth]{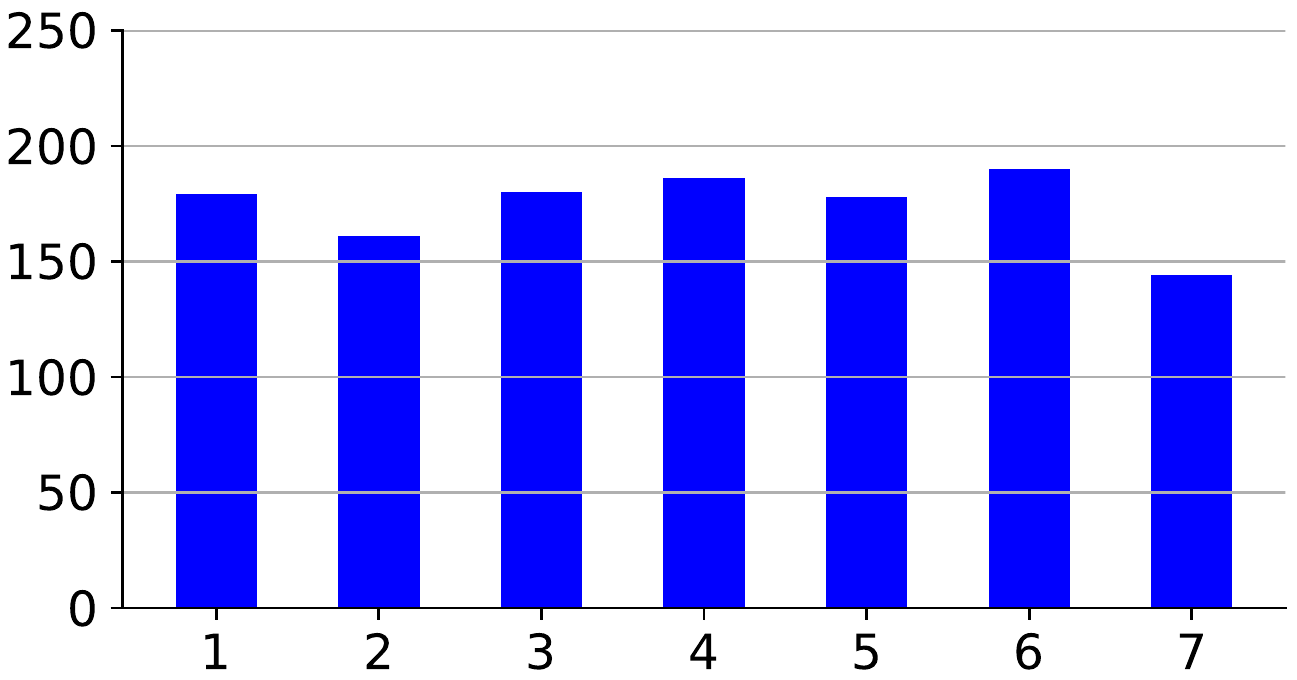}
		\end{minipage}\label{subfig:Time distribution5}
	}
	\subfigure[Filtered hour distribution of non-M/C]{
		\begin{minipage}[b]{0.31\textwidth}
			\includegraphics[width=1\textwidth]{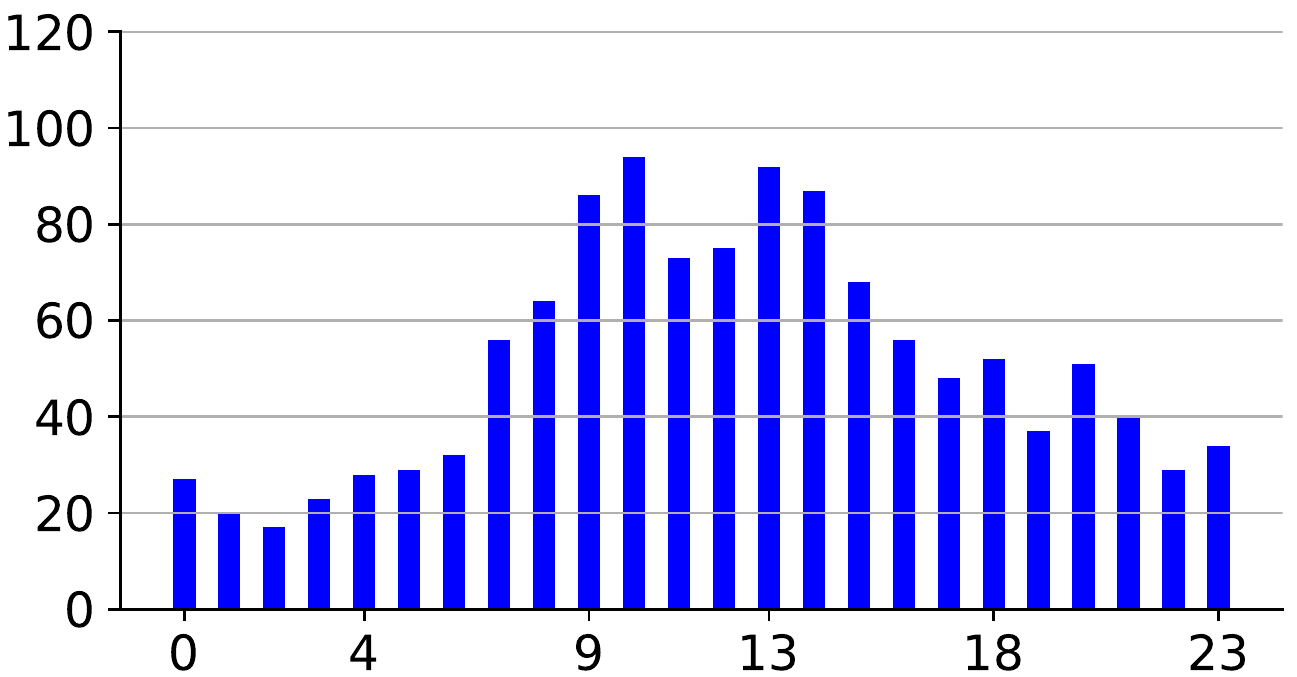}
		\end{minipage}\label{subfig:Time distribution6}
	}
	\caption{Original and filtered week, day and hour distribution of non-M/C. In each subfigure, the vertical axis is number of recordings and the horizontal axis is time intervals (52 weeks, 7 days and 24 hours).}\label{fig:Time distribution}
\end{figure*}

Fig. \ref{fig:Lolazation for nonMC} shows the location distribution of \textit{non-M/C}. 
It can be observed that there is only one outlier and the distance between it and others is about 20km.
Fig. \ref{subfig:Time distribution1}, \ref{subfig:Time distribution2} and \ref{subfig:Time distribution3} illustrate the original time distributions of week, day and hour for \textit{non-M/C}, which show significant imbalance on temporal scale.
Therefore, we filtered the location outlier and reduced the imbalance of time distribution.
By randomly reducing some recordings in the time categories with large number, the mean and variance values of week, day and hour distribution decrease, and training samples decreases from 1466 to 1218.
The filtered time distribution of \textit{non-M/C} are shown in Fig. \ref{subfig:Time distribution4}, \ref{subfig:Time distribution5} and \ref{subfig:Time distribution6}.
To further evaluate the filtering method, similar operations of time distribution were carried out for \textit{dog}.
The number of training samples for \textit{dog} decreases from 984 to 872.

Comparative results between single modality and multi-modality in Table \ref{tab:encoded mothod} show that the multi-modality performs worst than single modality for some classes, such as \textit{non-M/C} and \textit{dog}.
We propose to discard the outlier in location distribution and mitigate the imbalance of time distribution.
The temporal and spatial distributions are remained after applying the filtering method.
Multi-modality method outperforms single modality method, which demonstrates the effectiveness of incorporating spatiotemporal context into the UST system.
In addition, it also indicates that the collection of urban sound needs to be well designed and executed.

\subsubsection{Encoding methods}
AUPRC scores of coarse-level classes obtained by different encoding methods are shown in Table \ref{tab:encoded mothod}.
Compared with original vector, FC encoded vector achieves slightly better results on some classes.
Better results of LSTM encoded vector are obtained for some classes, such as \textit{M/C} and \textit{saw}.
Original encoded vector achieves good results for most of the classes, but shows disparity for \textit{dog}.

FC and LSTM methods can achieve better results on some classes, but these methods introduce more system complexities.
Conversely, simply fusing spatiotemporal vector with high-level audio representation also achieves good results.
This indicates that the method of encoding spatiotemporal context should be selected carefully.

\subsection{Comparisons in Each Module}
Regarding the modules in the multimodal UST system, comparisons from several aspects, i.e., features, CNNs with residual block, data augmentation and model fusion, are discussed.
All the results across sixteen models are given in Table \ref{tab:results}, and the best scores on coarse-level classes are highlighted.

\begin{table*}[t]
	\centering
	\setlength{\tabcolsep}{2.7mm}
	\renewcommand\arraystretch{1.25}
	\caption{Experimental results of models on each coarse-level class.}
	\begin{tabular}{ccccccccccc}
		\hline\hline
		Architecture                & Augmentation                 & Feature     & Engine                                & M/C                           & Non-M/C                        & Saw                           & Alert                         & Music                                 & Human                          & Dog                                   \\ \hline
		&                              & Log-Mel      & 0.877                                 & 0.637                                 & 0.586                                 & 0.758                                 & 0.948                                 & 0.682                                 & 0.976                                 & 0.306                                 \\
		&                              & Log-linear & 0.860                                 & 0.644                                 & 0.578                                 & 0.730                                 & 0.930                                 & 0.513                                 & 0.973                                 & 0.374                                 \\
		&                              & HPSS-h    & 0.869                                 & 0.618                                 & 0.501                                 & 0.720                                 & 0.939                                 & {\textbf{0.767}} & 0.974                                 & 0.266                                 \\
		& \multirow{-4}{*}{w/o mixup} & HPSS-p    & 0.872                                 & 0.677                                 & 0.595                                 & 0.743                                 & 0.888                                 & 0.478                                 & 0.976                                 & 0.380                                 \\ \cline{2-11}
		&                              & Log-Mel      & 0.874                                 & 0.586                                 & 0.557                                 & 0.787                                 & {\textbf{0.956}} & 0.658                                 & 0.972                                 & {\textbf{0.511}} \\
		&                              & Log-linear & 0.868                                 & 0.541                                 & 0.560                                 & 0.754                                 & 0.923                                 & 0.433                                 & 0.967                                 & 0.211                                 \\
		&                              & HPSS-h    & 0.865                                 & 0.601                                 & 0.491                                 & 0.725                                 & 0.930                                 & 0.658                                 & 0.966                                 & 0.188                                 \\
		\multirow{-8}{*}{CNN9}      & \multirow{-4}{*}{w/ mixup}  & HPSS-p    & 0.858                                 & 0.605                                 & 0.558                                 & 0.779                                 & 0.870                                 & 0.471                                 & 0.973                                 & 0.125                                 \\ \hline
		&                              & Log-Mel      & {\textbf{0.881}} & {\textbf{0.680}} & {\textbf{0.641}} & 0.738                                 & 0.945                                 & 0.729                                 & {\textbf{0.977}} & 0.361                                 \\
		&                              & Log-linear & 0.870                                 & 0.656                                 & 0.603                                 & 0.723                                 & 0.929                                 & 0.495                                 & 0.972                                 & 0.116                                 \\
		&                              & HPSS-h    & 0.879                                 & 0.616                                 & 0.531                                 & 0.702                                 & 0.946                                 & 0.748                                 & 0.974                                 & 0.149                                 \\
		& \multirow{-4}{*}{w/o mixup} & HPSS-p    & 0.867                                 & 0.609                                 & 0.609                                 & 0.739                                 & 0.889                                 & 0.519                                 & 0.976                                 & 0.258                                 \\ \cline{2-11}
		&                              & Log-Mel      & 0.876                                 & 0.602                                 & 0.607                                 & 0.755                                 & 0.943                                 & 0.678                                 & 0.972                                 & 0.304                                 \\
		&                              & Log-linear & 0.869                                 & 0.585                                 & 0.601                                 & 0.754                                 & 0.931                                 & 0.475                                 & 0.971                                 & 0.114                                 \\
		&                              & HPSS-h    & 0.873                                 & 0.649                                 & 0.481                                 & 0.742                                 & 0.931                                 & 0.722                                 & 0.968                                 & 0.150                                 \\
		\multirow{-8}{*}{CNN9-Res} & \multirow{-4}{*}{w/ mixup}  & HPSS-p    & 0.866                                 & 0.552                                 & 0.567                                 & {\textbf{0.788}} & 0.875                                 & 0.483                                 & 0.972                                 & 0.312                                 \\ \hline\hline
	\end{tabular}\label{tab:results}
\end{table*}

\subsubsection{Features}
\begin{figure}[!t]
	\centering
	\includegraphics[width=9cm,height=6cm]{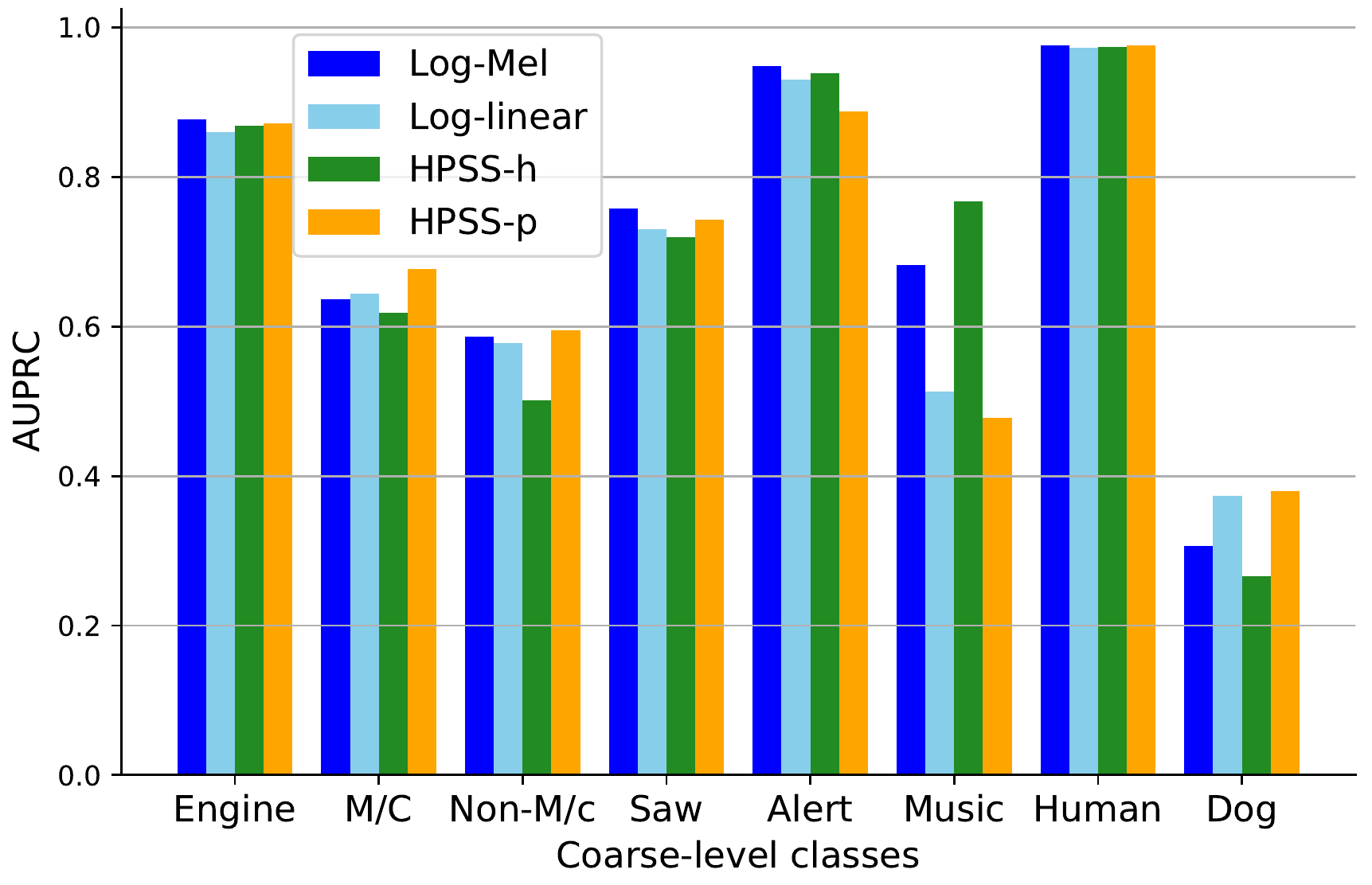}
	\caption{Comparison of different features based on CNN9 without mixup.}
	\label{fig:Comparisons1}
\end{figure}

We compared four features, i.e., log-Mel, log-linear, HPSS-h and HPSS-p spectrograms.
It is noted that the corresponding models are trained with CNN9.
The classification performance of different features is presented in Fig. \ref{fig:Comparisons1}, which clearly demonstrates the characteristic of each feature for tagging various urban sound.

In general, log-Mel spectrogram performs the best on coarse-level classes such as \textit{engine} and \textit{alert}, but its performance degrades for tagging \textit{dog}. 
Log-linear spectrogram gives good results on most of the coarse-level classes, especially \textit{dog}, but fails to yield good result for \textit{music}. 
HPSS-h spectrogram performs the best for \textit{music} detection, while for other classes it performs worse than other features. 
HPSS-p spectrogram is able to classify \textit{M/C}, but it fails to correctly classify \textit{music}.

Comparing the performance of features, it is observed that log-Mel spectrogram outperforms log-linear and HPSS spectrograms in general, and can recognize
most of the classes.
Since log-Mel spectrogram has high resolution in low frequencies, this indicates that these sounds in urban environment have discriminative characteristics in low frequencies.
In addition, HPSS-h spectrogram can correctly classify \textit{music}, which shows that the harmonic components in \textit{music} are more differentiable.

\subsubsection{CNN9 vs. CNN9-Res}

Experiments were carried out for comparing the performance of CNN9 and CNN9-Res. 
From Table \ref{tab:results}, it can be seen that CNN9-Res yields better results than CNN9 with different features.
This demonstrates that the residual block can effectively improve the performance of tagging urban sound.
Note that log-Mel spectrogram benefits the most from the residual block.

\subsubsection{Data augmentation}

In regard to the experiments of CNN9 using mixup, log-Mel spectrogram achieves the best AUPRC scores of 0.956 and 0.511 for classifying \textit{alert} and \textit{dog}, respectively.
Regarding the results of CNN9-Res using mixup, HPSS-h spectrogram achieves a higher score on \textit{M/C}, and HPSS-p spectrogram performs the best for classifying \textit{saw}.
Moreover, all the features can benefit from the data augmentation method for classifying \textit{saw}.

As it is shown in Fig. \ref{fig:At least and only one}, the data of \textit{alert} are not sufficient and the data of \textit{dog} and \textit{saw} are quite scarce.
For classifying \textit{dog}, applying mixup produces a significant improvement of 67\%.
The performance of \textit{alert} and \textit{saw} is also enhanced by mixup.
To summarize, data augmentation method can be applied to address the problem of lacking training data for specific urban sounds.

\subsubsection{Model Fusion}

\begin{figure}[t]
	\centering
	\includegraphics[width=9cm,height=6cm]{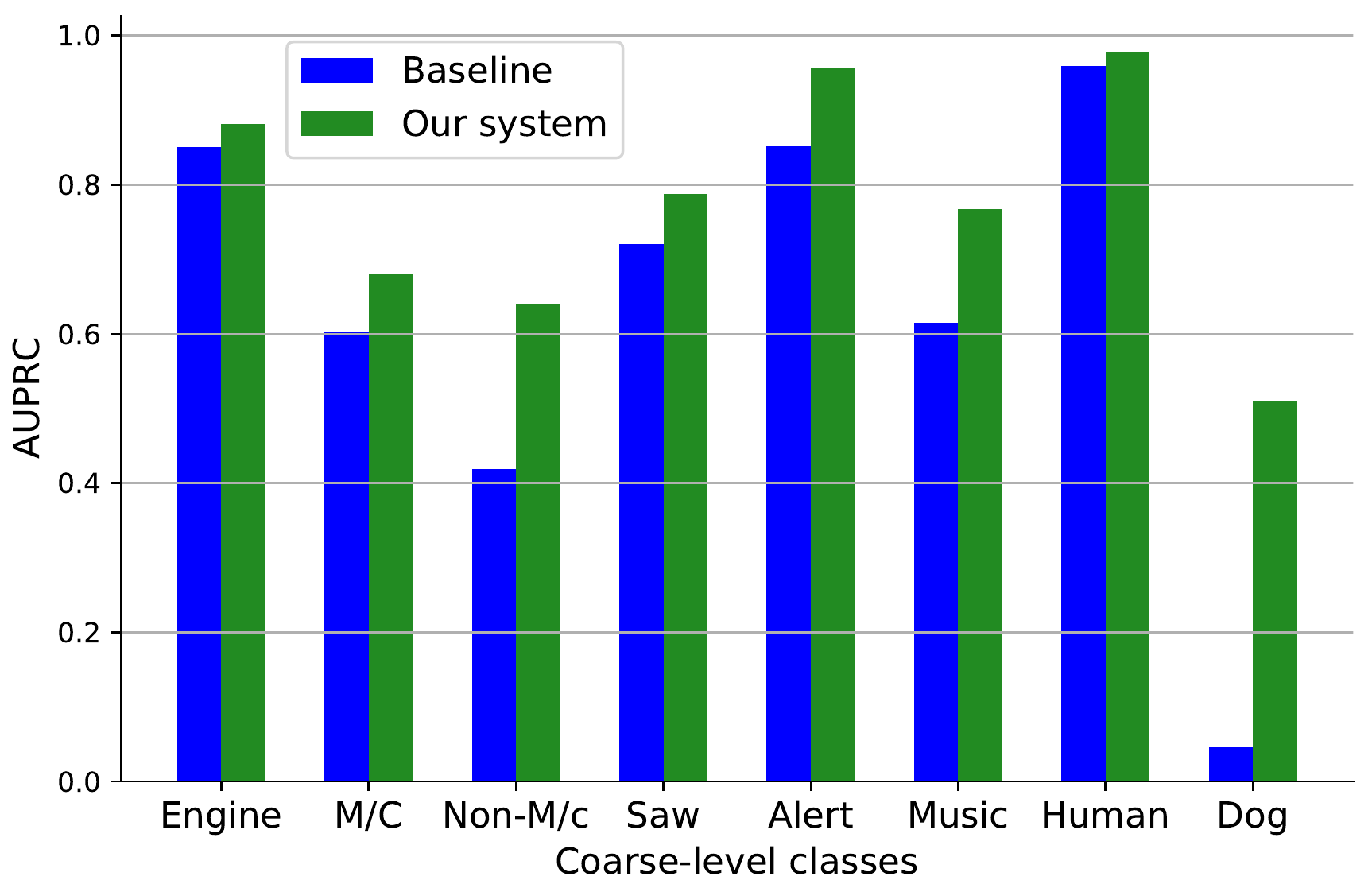}
	\caption{Comparison of the best AUPRC scores on each coarse-level class between our system and baseline system.}
	\label{fig:fusionvsbaseline}
\end{figure}

\begin{figure*}[t]
	\centering
	\includegraphics[width=14cm,height=7cm]{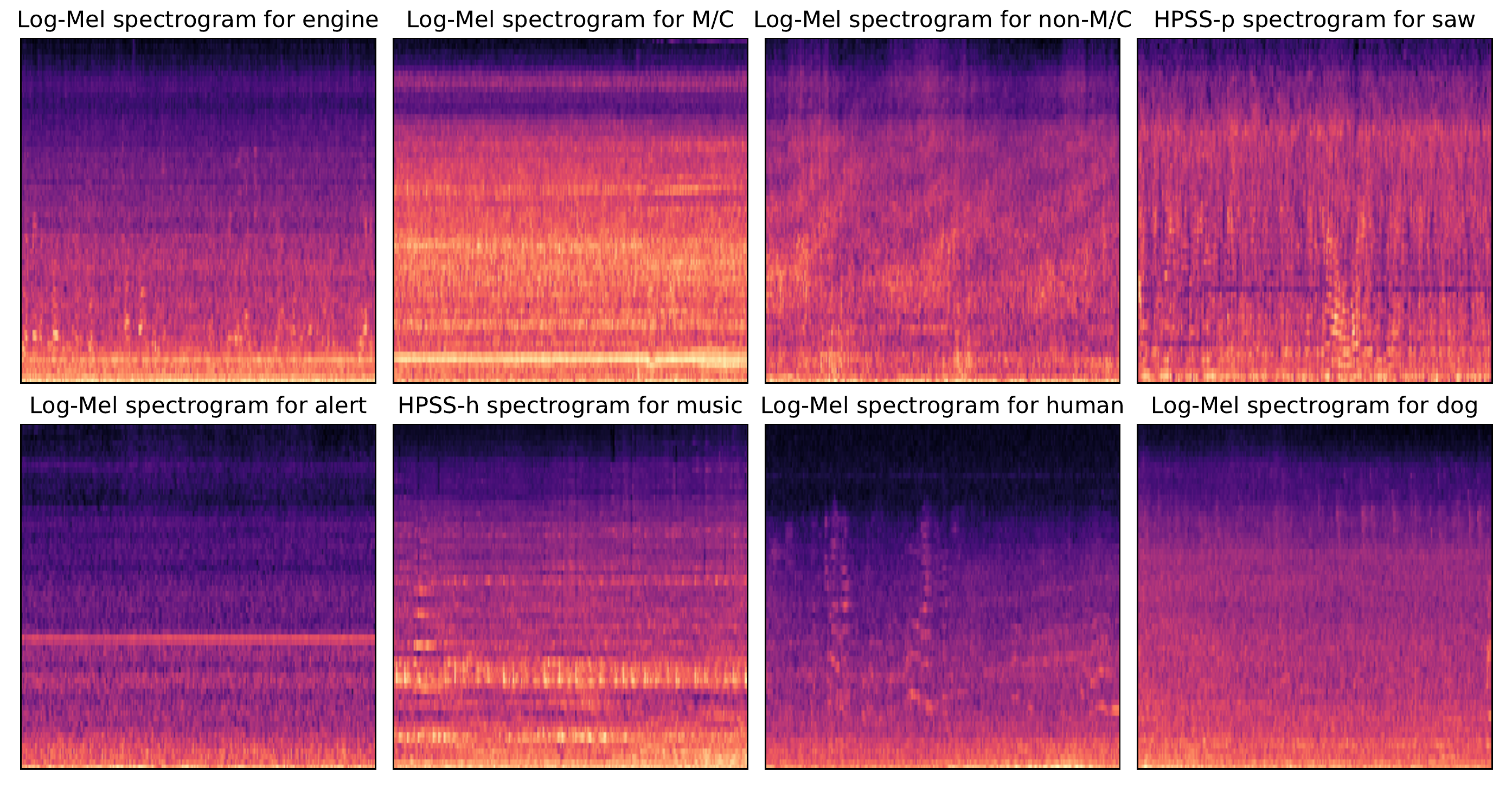}
	\caption{Examples of the spectrograms in coarse-level classes.}
	\label{fig:features examples}
\end{figure*}

\begin{table}[t]
	\centering
	\renewcommand\arraystretch{1.25}
	\caption{The best performance and its corresponding implementation on each coarse-level class.}
	\label{tab:best scores of methods}
	\begin{threeparttable}
	\begin{tabular}{ccccc}
		\hline\hline
		Coarse-level classes & AUPRC & Architecture      & Aug\tnote{*} & Feature \\ \hline
		Engine         & 0.881  & CNN9-Res & -            & Log-Mel \\
		M/C            & 0.680  & CNN9-Res & -            & Log-Mel \\
		Non-M/C        & 0.641  & CNN9-Res & -            & Log-Mel \\
		Saw            & 0.788  & CNN9-Res & Mixup       & HPSS-p  \\
		Alert          & 0.956  & CNN9     & Mixup       & Log-Mel \\
		Music          & 0.767  & CNN9     & -            & HPSS-h  \\
		Human          & 0.977  & CNN9-Res & -            & Log-Mel \\
		Dog            & 0.511  & CNN9     & Mixup       & Log-Mel \\ \hline\hline
	\end{tabular}
	\begin{tablenotes}
	\footnotesize
	\item[*] The term Aug is short for data augmentation. 
	\end{tablenotes}
	\end{threeparttable}

\end{table}
We take the Baseline-2020 as our baseline system.
Fig. \ref{fig:fusionvsbaseline} shows the class-wise AUPRC scores of baseline and proposed system, which outperforms the baseline on each coarse-level class. 
The AUPRC score of \textit{dog} is improved the most by 1010.87\%, and the AUPRC score of \textit{human} is improved the least by 1.88\%.
The models which achieve the best performance on coarse-level classes in Table \ref{tab:results} are summarized in Table \ref{tab:best scores of methods}.
Corresponding spectrograms of each coarse level class are shown in Fig. \ref{fig:features examples}.

To further boost the performance, an ensemble method was implemented to fuse the models. 
Finally, a macro-auprc score of 0.775 is achieved, which is nearly 22.63\% greater than the baseline system.

\subsection{Discrimination Analysis}

Keeping in view the complexities of urban environments, we further analyze the discrimination on coarse-level classes.
The study of confusion between classes can help us analyze the characteristic of urban sound and adjust the procedures of recording.
In the UST system, each recording in validate dataset can be annotated as multi-label.
For a given recording, the true label can be expressed as
$l_{v}=\left( l_{v,1}, l_{v,2}, \cdots, l_{v,n}\right) , n=8$, and the prediction which is sigmoid activated can be expressed as
$z_{v}=\left( z_{v,1}, z_{v,2}, \cdots, z_{v,n}\right) , n=8$.
For a pair of $l_{v}$ and $z_{v}$, we assume that if:
\begin{equation}
l_{v,i}=1 \, \& \, l_{v,j}=0 \, \& \, z_{v,j}=1 \, (\forall i\neq j,\, i,j \in n),
\end{equation}
the distracted class $j$ will be recorded.
The number and ratio of distractors are shown in Table \ref{tab:Discrimination analysis}.

\begin{table}[t]
	\centering
	\setlength{\tabcolsep}{4mm}
	\renewcommand\arraystretch{1.25}
	\caption{Discrimination analysis.}
	\label{tab:Discrimination analysis}
	\begin{tabular}{ccccc}
		\hline\hline
		\multicolumn{2}{c}{Single-label ($l_{v,i}$)}               & \multicolumn{3}{c}{Distractor ($z_{v,j}$)}       \\
		\cmidrule(lr){1-2} \cmidrule(lr){3-5}
		Classes          & Number               & Classes & Number & Ratio  \\ \hline
		\multirow{3}{*}{Engine} & \multirow{3}{*}{73}  & Human              & 4      & 4/73   \\
		&                      & Saw                & 2      & 2/73   \\
		&                      & Alert              & 1      & 1/73   \\ \hline
		\multirow{3}{*}{M/C}    & \multirow{3}{*}{11}  & Engine                & 3      & 3/11   \\
		&							& Saw             & 3      & 3/11   \\
		&                      & Human              & 1      & 1/11   \\\hline
		Non-M/C                 & 5                    & Engine             & 1      & 1/5    \\\hline
		Saw                     & 10                   & Engine             & 2     & 2/10   \\\hline
		\multirow{2}{*}{Alert}  & \multirow{2}{*}{37}  & Engine             & 10      & 10/37   \\
		&                      & Human              & 5      & 5/37   \\\hline
		\multirow{5}{*}{Music}  & \multirow{5}{*}{9}   & Human              & 6      & 6/9    \\
		&						& Engine             & 3      & 3/9    \\
		&						& Non-M/C              & 1      & 1/9    \\
		&						& Saw            & 1      & 1/9    \\
		&                      & Alert           & 1      & 1/9    \\\hline
		\multirow{5}{*}{Human}  & \multirow{5}{*}{104} & Engine             & 11     & 11/104 \\
		&                      & Alert              & 7      & 7/104  \\
		&                      & Music              & 2      & 2/104  \\
		&                      & Dog                & 2      & 2/104  \\
		&                      & Saw               & 1      & 1/104  \\\hline
		Dog                     & 2                    & Human                &1     & 1/2     \\ \hline\hline
	\end{tabular}
\end{table}

As shown in Table \ref{tab:Discrimination analysis},
\textit{engine} disrupts the detection of most classes, and it is the only distractor of \textit{non-M/C} and \textit{saw}.
The reason for this can be due to class-imbalance, i.e., the occurrence of \textit{engine} sound is more as compared with other sounds in the dataset.
We can find that \textit{engine} usually distracts the classification of mechanical sound due to the characteristics of \textit{engine}, \textit{saw} and other machine sounds are indistinct, which leads to the inaccurate labels.

Considering \textit{human}, we find \textit{human} and \textit{music} interfere with each other, and \textit{human} is the dominant distractor.
\textit{Music} in daily life contains not only instrument sounds but also a lot of vocals.
This leads to the confusion when tagging \textit{music}.
To ensure \textit{music} to be correctly classified, it is essential to learn the relationship between musical instruments and human voices for the classifiers.
Another observation is that confusion occurs in tagging \textit{alert} and \textit{human}. 
In some emergency cases, there could be \textit{alert} accompanied with human voices, e.g., shout or scream.
This is the reason why \textit{alert} and \textit{human} interfere with each other during the detection.
In general, the production of urban sound is predominantly related to human activities.
This may explain the confusions between \textit{human} and other classes.

\section{CONCLUSIONS}\label{sec:CONCLUSIONS}
In this paper, we propose a multimodal UST system based on audio and text modalities.
In our approach, two sets of four acoustic features, i.e., log-Mel, log-linear and HPSS spectrograms, are extracted from the urban sound recordings, which aims to fully integrate the complementary information of different time-frequency representations.
Then, these spectrograms are fed into a residual CNN, model ensemble and data augmentation method are applied to improve the performance. 
To conduct multimodal learning with text, the spatiotemporal vector is extracted from the spatiotemporal context by normalization and neural-network-based encoders, and concatenated with the audio representation learned by CNN for audio tagging. 
To our knowledge, this is the first time that text modality is used for UST.

Experimental results show that the proposed multimodal learning system with audio signals and spatiotemporal context outperforms the single modality system with audio signals only. 
This indicates that the location and time distribution have strong effects on the performance of UST.
The results also show that the proposed UST system outperforms the Baseline-2020 by a large margin. 
Moreover, log-Mel spectrogram performs better than other features, which may be due to its high resolution in low frequencies. For HPSS spectrograms, the harmonic component separated from \textit{music} is effective for classification and the percussive component improves the performance of tagging \textit{saw}.
These empirical observations inspire us to take advantage of complementary features.

We further analyze the discriminative classes of UST.
Engine sound can not be correctly classified with mechanical sounds, which inspires us to extract distinguishable feature for classifying machine sounds.
Human voices interfere the classification of other urban sounds, which inspires us to find effective method to exclude the interference of human.



\ifCLASSOPTIONcaptionsoff
\newpage
\fi



\bibliographystyle{IEEEtran}
\bibliography{refers}

\begin{thebibliography}{10}
\providecommand{\url}[1]{#1}
\csname url@samestyle\endcsname
\providecommand{\newblock}{\relax}
\providecommand{\bibinfo}[2]{#2}
\providecommand{\BIBentrySTDinterwordspacing}{\spaceskip=0pt\relax}
\providecommand{\BIBentryALTinterwordstretchfactor}{4}
\providecommand{\BIBentryALTinterwordspacing}{\spaceskip=\fontdimen2\font plus
\BIBentryALTinterwordstretchfactor\fontdimen3\font minus
  \fontdimen4\font\relax}
\providecommand{\BIBforeignlanguage}[2]{{%
\expandafter\ifx\csname l@#1\endcsname\relax
\typeout{** WARNING: IEEEtran.bst: No hyphenation pattern has been}%
\typeout{** loaded for the language `#1'. Using the pattern for}%
\typeout{** the default language instead.}%
\else
\language=\csname l@#1\endcsname
\fi
#2}}
\providecommand{\BIBdecl}{\relax}
\BIBdecl

\bibitem{Bello2019sonyc}
J.~P. Bello, C.~Silva, O.~Nov, R.~L. Dubois, A.~Arora, J.~Salamon, C.~Mydlarz,
  and H.~Doraiswamy, ``Sonyc: A system for monitoring, analyzing, and
  mitigating urban noise pollution,'' \emph{Communications of the ACM},
  vol.~62, no.~2, pp. 68--77, Feb 2019.

\bibitem{stansfeld2003noise}
S.~A. Stansfeld and M.~P. Matheson, ``Noise pollution: non-auditory effects on
  health,'' \emph{British medical bulletin}, vol.~68, no.~1, pp. 243--257,
  2003.

\bibitem{cartwright2019sonyc}
M.~Cartwright, A.~E.~M. Mendez, J.~Cramer, V.~Lostanlen, G.~Dove, H.-H. Wu,
  J.~Salamon, O.~Nov, and J.~Bello, ``{SONYC} urban sound tagging
  ({SONYC-UST}): A multilabel dataset from an urban acoustic sensor network,''
  in \emph{Proceedings of the Workshop on Detection and Classification of
  Acoustic Scenes and Events (DCASE)}, October 2019, pp. 35--39.

\bibitem{2013Novel}
M.~C. Bell and F.~Galatioto, ``Novel wireless pervasive sensor network to
  improve the understanding of noise in street canyons,'' \emph{Applied
  Acoustics}, vol.~74, no.~1, pp. 169 -- 180, 2013.

\bibitem{BaiWCCF19}
J.~Bai, B.~Wang, C.~Chen, J.~Chen, and Z.-H. Fu,
  ``\BIBforeignlanguage{English}{Inception-v3 based method of lifeclef 2019
  bird recognition},'' vol. 2380, Lugano, Switzerland, 2019.

\bibitem{salamon2015unsupervised}
J.~{Salamon} and J.~P. {Bello}, ``Unsupervised feature learning for urban sound
  classification,'' in \emph{2015 IEEE International Conference on Acoustics,
  Speech and Signal Processing (ICASSP)}, 2015, pp. 171--175.

\bibitem{zhang2020learning}
Z.~Zhang, S.~Xu, S.~Zhang, T.~Qiao, and S.~Cao, ``Learning frame level
  attention for environmental sound classification,'' \emph{arXiv preprint
  arXiv:2007.07241}, 2020.

\bibitem{adapa2019urban}
S.~Adapa, ``Urban sound tagging using convolutional neural networks,''
  \emph{arXiv preprint arXiv:1909.12699}, 2019.

\bibitem{Kim2019}
B.~Kim, ``Convolutional neural networks with transfer learning for urban sound
  tagging,'' DCASE2019 Challenge, Tech. Rep., September 2019.

\bibitem{kong2019cross}
Q.~Kong, Y.~Cao, T.~Iqbal, Y.~Xu, W.~Wang, and M.~D. Plumbley, ``Cross-task
  learning for audio tagging, sound event detection and spatial localization:
  Dcase 2019 baseline systems,'' \emph{arXiv preprint arXiv:1904.03476}, 2019.

\bibitem{Iqbal2020}
T.~Iqbal, Y.~Cao, M.~D. Plumbley, and W.~Wang, ``Incorporating auxiliary data
  for urban sound tagging,'' DCASE2020 Challenge, Tech. Rep., October 2020.

\bibitem{bai2019multi}
J.~{Bai}, C.~{Chen}, and J.~{Chen}, ``A multi-feature fusion based method for
  urban sound tagging,'' in \emph{2019 Asia-Pacific Signal and Information
  Processing Association Annual Summit and Conference (APSIPA ASC)}, 2019, pp.
  1313--1317.

\bibitem{he2016deep}
K.~He, X.~Zhang, S.~Ren, and J.~Sun, ``Deep residual learning for image
  recognition,'' in \emph{Proceedings of the IEEE conference on computer vision
  and pattern recognition}, 2016, pp. 770--778.

\bibitem{zhang2018deep}
Z.~Zhang, S.~Xu, S.~Cao, and S.~Zhang, ``\BIBforeignlanguage{English}{Deep
  convolutional neural network with mixup for environmental sound
  classification},'' vol. 11257 LNCS, Guangzhou, China, 2018, pp. 356 -- 367.

\bibitem{mesaros2016tut}
A.~Mesaros, T.~Heittola, and T.~Virtanen, ``Tut database for acoustic scene
  classification and sound event detection,'' in \emph{2016 24th European
  Signal Processing Conference (EUSIPCO)}.\hskip 1em plus 0.5em minus
  0.4em\relax IEEE, 2016, pp. 1128--1132.

\bibitem{chu2009environmental}
S.~Chu, S.~Narayanan, and C.-C.~J. Kuo, ``Environmental sound recognition with
  time--frequency audio features,'' \emph{IEEE Transactions on Audio, Speech,
  and Language Processing}, vol.~17, no.~6, pp. 1142--1158, 2009.

\bibitem{hershey2017cnn}
S.~Hershey, S.~Chaudhuri, D.~P. Ellis, J.~F. Gemmeke, A.~Jansen, R.~C. Moore,
  M.~Plakal, D.~Platt, R.~A. Saurous, B.~Seybold \emph{et~al.}, ``Cnn
  architectures for large-scale audio classification,'' in \emph{2017 ieee
  international conference on acoustics, speech and signal processing
  (icassp)}.\hskip 1em plus 0.5em minus 0.4em\relax IEEE, 2017, pp. 131--135.

\bibitem{Xu2017}
Y.~Xu, Q.~Kong, W.~Wang, and M.~D. Plumbley, ``Surrey-{CVSSP} system for
  {DCASE2017} challenge task4,'' DCASE2017 Challenge, Tech. Rep., September
  2017.

\bibitem{vuegen2018weakly}
L.~Vuegen, P.~Karsmakers, B.~Vanrumste \emph{et~al.}, ``Weakly-supervised
  classification of domestic acoustic events for indoor monitoring
  applications,'' in \emph{In proceedings of IEEE Conference on Biomedical and
  Health Informatics 2018}.\hskip 1em plus 0.5em minus 0.4em\relax IEEE, 2018.

\bibitem{cowling2003comparison}
M.~Cowling and R.~Sitte, ``Comparison of techniques for environmental sound
  recognition,'' \emph{Pattern recognition letters}, vol.~24, no.~15, pp.
  2895--2907, 2003.

\bibitem{heittola2013context}
T.~Heittola, A.~Mesaros, A.~Eronen, and T.~Virtanen, ``Context-dependent sound
  event detection,'' \emph{EURASIP Journal on Audio, Speech, and Music
  Processing}, vol. 2013, no.~1, p.~1, 2013.

\bibitem{salamon2014dataset}
J.~Salamon, C.~Jacoby, and J.~P. Bello, ``A dataset and taxonomy for urban
  sound research,'' in \emph{Proceedings of the 22nd ACM international
  conference on Multimedia}, 2014, pp. 1041--1044.

\bibitem{cotton2011spectral}
C.~V. Cotton and D.~P. Ellis, ``Spectral vs. spectro-temporal features for
  acoustic event detection,'' in \emph{2011 IEEE Workshop on Applications of
  Signal Processing to Audio and Acoustics (WASPAA)}.\hskip 1em plus 0.5em
  minus 0.4em\relax IEEE, 2011, pp. 69--72.

\bibitem{serizel2018large}
R.~Serizel, N.~Turpault, H.~Eghbal-Zadeh, and A.~P. Shah, ``Large-scale weakly
  labeled semi-supervised sound event detection in domestic environments,''
  \emph{arXiv preprint arXiv:1807.10501}, 2018.

\bibitem{salamon2017deep}
J.~{Salamon} and J.~P. {Bello}, ``Deep convolutional neural networks and data
  augmentation for environmental sound classification,'' \emph{IEEE Signal
  Processing Letters}, vol.~24, no.~3, pp. 279--283, 2017.

\bibitem{cakir2017convolutional}
E.~Cak{\i}r, G.~Parascandolo, T.~Heittola, H.~Huttunen, and T.~Virtanen,
  ``Convolutional recurrent neural networks for polyphonic sound event
  detection,'' \emph{IEEE/ACM Transactions on Audio, Speech, and Language
  Processing}, vol.~25, no.~6, pp. 1291--1303, 2017.

\bibitem{akiyama2019multitask}
O.~Akiyama and J.~Sato, ``Dcase 2019 task 2: Multitask learning,
  semi-supervised learning and model ensemble with noisy data for audio
  tagging,'' 01 2019, pp. 25--29.

\bibitem{ngiam2011multimodal}
J.~Ngiam, A.~Khosla, M.~Kim, J.~Nam, H.~Lee, and A.~Ng, ``Multimodal deep
  learning,'' 01 2011, pp. 689--696.

\bibitem{srivastava2014multimodal}
N.~Srivastava and R.~Salakhutdinov, ``Multimodal learning with deep boltzmann
  machines,'' \emph{The Journal of Machine Learning Research}, vol.~15, no.~1,
  pp. 2949--2980, 2014.

\bibitem{Zambellionline}
M.~{Zambelli} and Y.~{Demirisy}, ``Online multimodal ensemble learning using
  self-learned sensorimotor representations,'' \emph{IEEE Transactions on
  Cognitive and Developmental Systems}, vol.~9, no.~2, pp. 113--126, 2017.

\bibitem{ramachandram2017deep}
D.~Ramachandram and G.~W. Taylor, ``Deep multimodal learning: A survey on
  recent advances and trends,'' \emph{IEEE Signal Processing Magazine},
  vol.~34, no.~6, pp. 96--108, 2017.

\bibitem{liu2018active}
H.~Liu, F.~Wang, F.~Sun, and X.~Zhang, ``Active visual-tactile cross-modal
  matching,'' \emph{IEEE Transactions on Cognitive and Developmental Systems},
  vol.~11, no.~2, pp. 176--187, 2018.

\bibitem{jan2017artificial}
A.~Jan, H.~Meng, Y.~F. B.~A. Gaus, and F.~Zhang, ``Artificial intelligent
  system for automatic depression level analysis through visual and vocal
  expressions,'' \emph{IEEE Transactions on Cognitive and Developmental
  Systems}, vol.~10, no.~3, pp. 668--680, 2017.

\bibitem{ringeval2013introducing}
F.~Ringeval, A.~Sonderegger, J.~Sauer, and D.~Lalanne, ``Introducing the recola
  multimodal corpus of remote collaborative and affective interactions,'' in
  \emph{2013 10th IEEE international conference and workshops on automatic face
  and gesture recognition (FG)}.\hskip 1em plus 0.5em minus 0.4em\relax IEEE,
  2013, pp. 1--8.

\bibitem{poria2016fusing}
S.~Poria, E.~Cambria, N.~Howard, G.-B. Huang, and A.~Hussain, ``Fusing audio,
  visual and textual clues for sentiment analysis from multimodal content,''
  \emph{Neurocomputing}, vol. 174, pp. 50--59, 2016.

\bibitem{simonyan2014two}
K.~Simonyan and A.~Zisserman, ``Two-stream convolutional networks for action
  recognition in videos,'' in \emph{Advances in neural information processing
  systems}, 2014, pp. 568--576.

\bibitem{yang2017eeg}
Y.~Yang, Q.~J. Wu, W.-L. Zheng, and B.-L. Lu, ``Eeg-based emotion recognition
  using hierarchical network with subnetwork nodes,'' \emph{IEEE Transactions
  on Cognitive and Developmental Systems}, vol.~10, no.~2, pp. 408--419, 2017.

\bibitem{kathania2019role}
H.~K. Kathania, S.~Shahnawazuddin, W.~Ahmad, and N.~Adiga, ``Role of linear,
  mel and inverse-mel filterbanks in automatic recognition of speech from
  high-pitched speakers,'' \emph{Circuits, Systems, and Signal Processing},
  vol.~38, no.~10, pp. 4667--4682, 2019.

\bibitem{huzaifah2017comparison}
M.~Huzaifah, ``Comparison of time-frequency representations for environmental
  sound classification using convolutional neural networks,'' \emph{arXiv
  preprint arXiv:1706.07156}, 2017.

\bibitem{fitzgerald2010harmonic}
D.~Fitzgerald, ``Harmonic/percussive separation using median filtering,''
  \emph{13th International Conference on Digital Audio Effects (DAFx-10)}, 01
  2010.

\bibitem{tachibana2013singing}
H.~Tachibana, N.~Ono, and S.~Sagayama, ``Singing voice enhancement in monaural
  music signals based on two-stage harmonic/percussive sound separation on
  multiple resolution spectrograms,'' \emph{IEEE/ACM Transactions on Audio,
  Speech, and Language Processing}, vol.~22, no.~1, pp. 228--237, 2013.

\bibitem{ioffe2015batch}
S.~Ioffe and C.~Szegedy, ``Batch normalization: Accelerating deep network
  training by reducing internal covariate shift,'' \emph{arXiv preprint
  arXiv:1502.03167}, 2015.

\bibitem{xu2015empirical}
B.~Xu, N.~Wang, T.~Chen, and M.~Li, ``Empirical evaluation of rectified
  activations in convolutional network,'' \emph{arXiv preprint
  arXiv:1505.00853}, 2015.

\bibitem{8434391}
B.~McFee, J.~Salamon, and J.~P. Bello, ``Adaptive pooling operators for weakly
  labeled sound event detection,'' \emph{IEEE/ACM Transactions on Audio,
  Speech, and Language Processing}, vol.~26, no.~11, pp. 2180--2193, 2018.

\bibitem{wang2019}
M.~{Wang}, R.~{Wang}, X.~{Zhang}, and S.~{Rahardja}, ``Hybrid constant-q
  transform based cnn ensemble for acoustic scene classification,'' in
  \emph{2019 Asia-Pacific Signal and Information Processing Association Annual
  Summit and Conference (APSIPA ASC)}, 2019, pp. 1511--1516.

\bibitem{cartwright2020sonyc}
M.~Cartwright, J.~Cramer, A.~E.~M. Mendez, Y.~Wang, H.-H. Wu, V.~Lostanlen,
  M.~Fuentes, G.~Dove, C.~Mydlarz, J.~Salamon \emph{et~al.}, ``Sonyc-ust-v2: An
  urban sound tagging dataset with spatiotemporal context,'' \emph{arXiv
  preprint arXiv:2009.05188}, 2020.

\bibitem{zhang2017mixup}
H.~Zhang, M.~Cisse, Y.~N. Dauphin, and D.~Lopez-Paz, ``mixup: Beyond empirical
  risk minimization,'' \emph{arXiv preprint arXiv:1710.09412}, 2017.

\bibitem{Diez2020}
I.~Diez, P.~Gonzalez, and I.~Gonzalez, ``Urban sound classification using
  convolutional neural networks for {DCASE} 2020 challenge,'' DCASE2020
  Challenge, Tech. Rep., October 2020.

\bibitem{Cartwright2020}
M.~Cartwright, J.~Cramer, A.~E. Mendez~Mendez, Y.~Wang, H.-H. Wu, V.~Lostanlen,
  M.~Fuentes, J.~Salamon, and J.~P. Bello, ``An urban sound tagging dataset
  with spatiotemporal context,'' DCASE2020 Challenge, Tech. Rep., October 2020.

\bibitem{8682475}
J.~Cramer, H.-H. Wu, J.~Salamon, and J.~P. Bello, ``Look, listen, and learn
  more: Design choices for deep audio embeddings,'' in \emph{ICASSP 2019 - 2019
  IEEE International Conference on Acoustics, Speech and Signal Processing
  (ICASSP)}, 2019, pp. 3852--3856.

\bibitem{dauphin2017language}
Y.~N. Dauphin, A.~Fan, M.~Auli, and D.~Grangier, ``Language modeling with gated
  convolutional networks,'' in \emph{International conference on machine
  learning}.\hskip 1em plus 0.5em minus 0.4em\relax PMLR, 2017, pp. 933--941.

\bibitem{park2019specaugment}
D.~S. Park, W.~Chan, Y.~Zhang, C.-C. Chiu, B.~Zoph, E.~D. Cubuk, and Q.~V. Le,
  ``Specaugment: A simple data augmentation method for automatic speech
  recognition,'' \emph{arXiv preprint arXiv:1904.08779}, 2019.

\bibitem{kong2020panns}
Q.~Kong, Y.~Cao, T.~Iqbal, Y.~Wang, W.~Wang, and M.~D. Plumbley, ``Panns:
  Large-scale pretrained audio neural networks for audio pattern recognition,''
  \emph{IEEE/ACM Transactions on Audio, Speech, and Language Processing},
  vol.~28, pp. 2880--2894, 2020.

\bibitem{gemmeke2017audio}
J.~F. Gemmeke, D.~P. Ellis, D.~Freedman, A.~Jansen, W.~Lawrence, R.~C. Moore,
  M.~Plakal, and M.~Ritter, ``Audio set: An ontology and human-labeled dataset
  for audio events,'' in \emph{2017 IEEE International Conference on Acoustics,
  Speech and Signal Processing (ICASSP)}.\hskip 1em plus 0.5em minus
  0.4em\relax IEEE, 2017, pp. 776--780.

\end{thebibliography}
%



%

\begin{IEEEbiography}[{\includegraphics[width=1in,height=1.25in,clip,keepaspectratio]{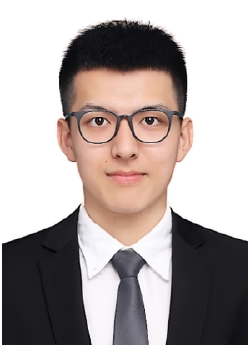}}]{Jisheng Bai}
received the B.S. degree in Detection Guidance and Control Technology from North University of China in 2017. 
He received the M.S. degree in Electronics and Communications Engineering from Northwestern Polytechnical University in 2020, where he is pursuing the Ph.D. degree. 
His research interests are focused on deep learning and environmental signal processing.
\end{IEEEbiography}

\begin{IEEEbiography}[{\includegraphics[width=1in,height=1.25in,clip,keepaspectratio]{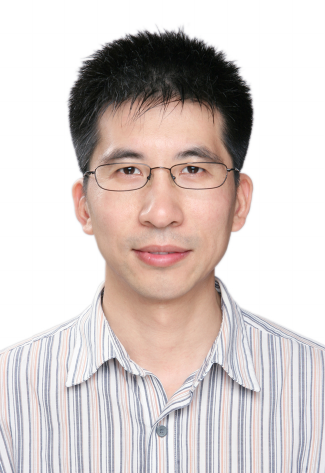}}]{Jianfeng Chen}
received the Ph.D. degree in 1999, from Northwestern Polytechnical University. 
From 1999 to 2001, he was a Research Fellow in School of EEE, Nanyang Technological University, Singapore.
During 2001-2003, he was with Center for Signal Processing, NSTB, Singapore, as a Research Scientist. 
From 2003 to 2007, he was a Research Scientist in Institute for Infocomm Research, Singapore. 
Since 2007, he works in College of Marine as a professor in Northwestern Polytechnical University, Xi’an, China. 
His main research interests include autonomous underwater vehicle design and application, array processing, acoustic signal processing, target detection and localization.
\end{IEEEbiography}

\begin{IEEEbiography}[{\includegraphics[width=1in,height=1.25in,clip,keepaspectratio]{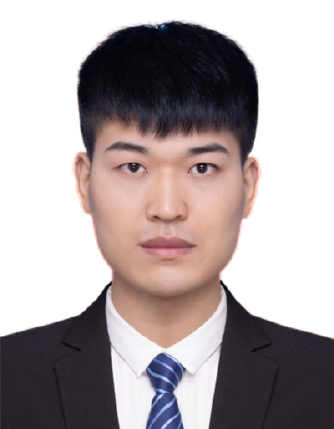}}]{Mou Wang}
(Student Member, IEEE) received the B.S. degree in electronics and information engineering from Northwestern Polytechnical University, China, in 2016, where he is pursuing the Ph.D. degree in information and communication engineering. His research interests include machine learning and speech signal processing. He was awarded Outstanding Reviewer of IEEE Transactions on Multimedia.
\end{IEEEbiography}







\end{document}